\documentclass[aps,amssymb,twocolumn,prb,superscriptaddress,showpacs]{revtex4}
\usepackage{bm}
\usepackage{graphicx}
\usepackage{color}

\begin{document}

\title{Quantum transport properties of two-dimensional electron gases under high magnetic fields}

\author{Thierry Champel}

\affiliation{
Institut f\"{u}r Theoretische Festk\"{o}rperphysik,
  Universit\"{a}t Karlsruhe,
  D-76128 Karlsruhe, Germany }
\affiliation{Laboratoire de Physique et Mod\'{e}lisation des Milieux Condens\'{e}s, CNRS et Universit\'{e} Joseph Fourier, 25 Avenue des Martyrs, BP 166, F-38042 Grenoble Cedex 9, France}
\author{Serge Florens}
\affiliation{
Institut f\"{u}r Theorie der Kondensierten Materie, Universit\"{a}t Karlsruhe, D-76128 Karlsruhe, Germany }
\affiliation{Institut N\'{e}el, CNRS et Universit\'{e} Joseph Fourier, 25 Avenue des Martyrs, BP 166, F-38042 Grenoble Cedex 9, France}
\date{\today}

\begin{abstract}
We study quantum transport properties of two-dimensional electron gases under high perpendicular magnetic fields. For this purpose, we reformulate the high-field expansion, usually done in the operatorial language of the guiding-center coordinates, in terms of vortex states within the framework of real-time Green functions. These vortex states arise naturally from the consideration that the Landau levels quantization can follow directly from the existence of a topological winding number.
The microscopic computation of the current  can then be performed within the Keldysh formalism
in a systematic way at finite magnetic fields $B$ (i.e. beyond the semi-classical limit $B = \infty$).
The formalism allows us to define a general vortex current density as long as the gradient expansion theory is applicable. As a result, the total current is expressed in terms of edge contributions only.
 We obtain  the first and third lowest order contributions to the current due to Landau-levels mixing processes, and derive in a transparent way the quantization of the Hall conductance. 
Finally, we point out qualitatively the importance of  inhomogeneities of the vortex density to capture the dissipative longitudinal transport.
\end{abstract}

\pacs{73.43.-f,72.15.Rn,73.50.Jt}

\maketitle

\section{Introduction}

As superfluidity and superconductivity, the integer quantum Hall effect is a spectacular manifestation of the quantum nature at a macroscopic scale.
Interestingly, these three phenomena share the same property of
disappearance of the dissipation at low temperatures.
In the two-dimensional electron gases in  the regime of the quantum Hall effect, the vanishing of the longitudinal magnetoresistance $R_{xx}$ for a range in magnetic fields, is accompanied by another remarkable feature which is the formation of plateaus of the Hall conductance $G_{xy}$ with a robust quantization \cite{vonK} in integral multiples of $e^{2}/h$.

There is now a consensus \cite{Prange,Hadju,Huckestein} that the existence of $G_{xy}$ plateaus is intimately connected with the localization mechanism which causes the vanishing of dissipation.
In a  phenomenological picture the  localized states that do not contribute to the longitudinal transport play the role of a reservoir for electrons which allows the chemical potential to sit between Landau levels $\varepsilon_{n}=(n+1/2) \hbar \omega_{c}$ (here $\omega_{c}$ is the cyclotron pulsation).
On the other hand, delocalized states are supposed to exist only within small energy bands centered around  the energies $\varepsilon_{n}$. As long as the chemical potential lies within the region of localized states, the Hall conductance $G_{xy}$ is expected to remain constant and is related to the number of filled bands of delocalized states below the chemical potential.
The observation of wide plateaus implies consequently a large density for the localized states with a much more significant spread in energy than for the density of delocalized states.
The density of states with a mobility gap is therefore the essential condition for the quantum Hall effect. \cite{Prange}


Many different theoretical approaches \cite{Laughlin,Halperin,Streda,Thouless,Prange2,Iordanski,Trugman,Kazarinov,Joynt,Apenko1985,Shapiro,MacDonald,Buttiker} have been proposed to explain the quantum Hall effect, which can be divided
in several classes. The most fruitful treatments, such as Laughlin's gauge argument, \cite{Laughlin} and the
edge state picture, \cite{Halperin,MacDonald,Buttiker} provide an explanation for the quantization of the Hall conductance, but rely on the assumption that there exist localized states in the bulk. 
Direct diagrammatic calculations have been developed as well. Models with point-like impurities are technically manageable, \cite{Streda} but require the artificial pinning of the chemical potential within gaps in the density of states to account for the existence of quantized Hall plateaus. Further developments to include finite-range disorder within the Green functions diagrammatics were pursued, but appear more complicated,
since it is needed to incorporate more and more diagrams with increasing strength of the magnetic field. \cite{Rai}

It is worth mentioning that the standard procedure \cite{Abrikosov} that consists in
averaging over impurity positions is in fact questionable in high magnetic fields. Indeed, it is normally justified by the physical assumption of randomness after successive collision events. However, instead of the chaotic random walk,  the electronic motion is relatively regular for high magnetic fields and smooth disorder potential. At the classical level, this leads to the breakdown of the Drude-Lorentz formula. \cite{Fog} Quantum mechanically, this idea has led to the most simple and illuminating approach to explain the existence of Hall plateaus, the guiding-center picture, \cite{Hadju,Iordanski,Trugman,Kazarinov,Joynt,Apenko1983,Apenko1984,Apenko1985,Shapiro} which is a high-field approximation. Within this "semi-classical" description,
the quantized cyclotron motion energy becomes a constant of motion when the disorder potential is smooth and varies less than the Landau levels separation (in order to have negligible Landau levels mixing).
The reason for this is an ineffective energy exchange between the two degrees of freedom, the fast cyclotron motion, and the slow guiding center motion. Without such an exchange, the strong field localization is then associated with the drift motion of the guiding center of the cyclotron orbit along contours of constant disorder potential.
The delocalization of electronic states is then due to the percolation of equipotential lines, that occurs at a single critical energy.
The localization is thus determined in terms of the topology of the equipotential lines, and the resulting spectral properties vindicate the mobility gap argument for the appearance of the quantum Hall effect in high fields. 

The computation of quantum transport beyond the $B=\infty$ limit has however not been achieved in a systematic way.
\cite{Hadju,Trugman,Kazarinov,Apenko1985,Shapiro} Because the matrix elements of the current density relate adjacent Landau levels, the mixing of Landau levels (which is absent at infinite magnetic field) needs thus to be taken into
account in a rigorous way. In this paper, we reformulate the high field expansion in the framework of real-time Green functions using a particular set of vortex eigenstates.
The framework developed in our paper, although formally equivalent to the guiding center approach, allows a straightforward and systematic calculation of quantum transport at finite magnetic fields $B$ (i.e. beyond the semi-classical limit $B= \infty$), and leads to a transparent derivation of both the dissipationless longitudinal conductance and the quantized Hall resistance.
We also show starting from our bulk calculation that the conductance can be expressed in terms of edge contributions, thus providing a connection between the high-field approaches and the edge state picture. This result is found up to the third order in processes mixing the Landau-levels, and may be a general result valid at all orders as long as the gradient expansion theory is applicable.
We interpret this as a fundamental manifestation of the robustness of the electronic vortex states, which gives rise to strong localization properties at large but finite magnetic fields and in long-range disorder.

The outline of the paper is as follows.
In Sec. II, we consider the simple problem of a single charged particle in a magnetic field.
We
first analyze within a gauge-independent framework the origin of the Landau levels energy quantization in terms of the wave-particle duality.
Such a preliminary discussion is needed because this origin is usually obscured by the choice of a gauge and the high degeneracy of the energy levels. We derive a basis of eigenstates in a way  which shows that the energy quantization follows {\it only} from the single-valuedness of the wavefunction, i.e. from a topological condition.
In Sec. III, by using this basis of vortex states, we
introduce within the framework of the real-time Green functions the high-field expansion that allows to treat a smooth disorder. In Sec. IV we study thermodynamic properties, and especially compute the dependence of the chemical potential on the magnetic field, that is important to determine the width of the Hall plateaus for different temperatures and disorder broadenings. Quantum transport properties in high magnetic fields are investigated in Sec. V.
 We discuss and summarize our work in Section VI. Some of the technical details such as the derivation and the resolution of the kinetic equation have been collected in the appendices.

\section{PRELIMINARIES}

In this Section, we consider a single electron of effective mass $m^{\ast}$ and
of charge $e=-|e|$ confined in a $(xy)$ two-dimensional  plane under a
perpendicular magnetic field ${\bf B}$ (pointing in the $z$ direction). The
influence of an impurity potential is postponed to Section III. The hamiltonian has the form

  \begin{eqnarray}
    H_{0}=\left({\bf p}-\frac{e}{c} {\bf A}\right)^{2}/2m^{\ast}=
\left(-i \hbar \bm{\nabla}+\frac{|e|}{c} {\bf A}\right)^{2}/2m^{\ast}
  \end{eqnarray}
where  ${\bf A}$ is the vector potential which is determined from the equation
\begin{equation}
 \bm{\nabla} \times {\bf A} = {\bf B}=B \hat{{\bf z}}.
\end{equation}
The vector potential is defined by this way up to the gradient of an arbitrary function  $\Lambda({\bf r})$ (i.e. a gauge factor).
Here, for simplicity, we do not consider the spin degree of freedom and disregard therefore the Zeeman splitting term.
The wave functions  $\Psi$ and the energy eigenvalues $E$ are derived from the stationary Schr\"{o}dinger's equation

\begin{eqnarray}
    H_{0} \Psi =E \Psi \label{VP}.
 \end{eqnarray}

Because it is the vector potential ${\bf A}$ rather than the magnetic field ${\bf B}$ which appears in $H_{0}$, there is some freedom in solving Eq. ($\ref{VP}$). In the next two subsections we introduce the two most known ways presented in textbooks to solve it, before analyzing the question of the origin of the quantization of the energy spectrum.

\subsection{Translation-invariant eigenstates} \label{Sec_trans}

The traditional way \cite{Landau} to solve the eigen-problem (\ref{VP}) is to introduce the Landau gauge, for which the vector potential has the form
 ${\bf A}=xB \hat{{\bf y}}$.  The hamiltonian is then also obviously invariant by translation in the $y$ direction so that the wave function is sought under the form
 $$\Psi(x,y)=f_{p}(x) \exp(ip y)$$ with $p$ real. The function $f_{p}(x)$ obeys the differential equation

\begin{equation}
 f''_{p}(x)+\left(\frac{2m^{\ast}E}{\hbar^{2}}-\left(p+ \frac{x }{l_{B}^{2}}\right)^{2}\right)f_{p}
= 0 \label{eq1}
\end{equation}
typical of an harmonic oscillator.
Here we have introduced the magnetic length
\begin{equation}
l_{B}=\sqrt{\hbar c/|e| B},
\end{equation}
which is related to the quantum of magnetic flux $\Phi_{0}$ as follows
\begin{equation}
\Phi_{0}=2 \pi l_{B}^{2} B = \frac{hc}{|e|}.
\end{equation}

Introducing the new coordinate $x'=x-x_{0}$ with  $x_{0}=-p l_{B}^{2}$, a solution of (\ref{eq1}) is sought under the form
$$f_{p}(x)= \exp \left[-\frac{x'^{2}}{2 l_{B}^{2}}\right] h(x')
$$
where $h(x')$ is expanded in power series.
The condition that $f_{p} \to 0$ for $x \to \infty$ imposes that the function $h$ is a polynomial of order $n$ and
leads to the well-known result of the quantization of the energy spectrum
\begin{equation}
E_{n,p}=\hbar \omega_{c}\left(
n+\frac{1}{2}
\right), \label{qu1}
\end{equation}
where $\omega_{c}=|e|B/m^{\ast}c$ is the cyclotron pulsation.
Here the energy $E$ is independent of the quantum number
 $p$. This expresses an important degeneracy of the energy levels. Finally, the wave functions are thus labeled by the quantum numbers
 $n$ and $p$ and are written as
\begin{equation}
\Psi_{n,p}(x,y)=e^{ipy} \exp \left[-\frac{(x-x_{0})^{2}}{2 l_{B}^{2}}\right] H_{n}\left(\frac{x-x_{0}}{l_{B}}\right) \label{trans}
\end{equation}
with  $H_{n}$ the Hermite polynomial of degree $n$.
Within this basis, the moduli $|\Psi_{n,p}(x,y)|$ have the property of being translation invariant in the $y$ direction, and the eigenstates $\Psi_{n,p}(x,y)$ are not square-integrable in the plane.

\subsection{Rotation-invariant eigenstates} \label{Sec_rot}

Instead of the Landau gauge, one can use the symmetrical gauge for which the vector potential is ${\bf A}= {\bf B} \times {\bf r}/2$. Introducing the polar coordinates $(r, \theta)$, one can straightforwardly see  that the hamiltonian $H_{0}$ is invariant by any rotation around the origin. Then a convenient form for the wave function is

\begin{equation}
\Psi(r,\theta)=f_{m}(r)\, \exp(i m \theta)
\end{equation}
with $m$ a positive or negative integer, condition required by the single-valuedness of the wave function. The function $f_{m}(r)$ obeys the one-dimensional differential equation

\begin{equation}
\frac{1}{r}
\frac{d}{d r} \left(r \frac{d f_{m}}{d r}\right)+\left(\frac{2m^{\ast}E}{\hbar^{2}}-\left(\frac{m}{r}+
\frac{r}{2 l_{B}^{2}}\right)^{2}\right)f_{m}
= 0. \label{sim}
\end{equation}
We search the function $f_{m}$ under the form

\begin{equation}
f_{m}(r)=r^{|m|} \, e^{-r^{2}/4 l_{B}^{2}} \,g_{m}(r).
\end{equation}
The function  $g_{m}$ is then given by the equation

\begin{eqnarray}
\frac{d^{2}g_{m}}{d r^{2}}
+\left(\frac{2 |m|+1}{r}-\frac{r}{l_{B}^{2}}
\right) \frac{d g_{m}}{dr}
\nonumber
\\+\frac{2m^{\ast}}{\hbar^{2}}
\left[E-
\frac{\hbar \omega_{c}}{2}\left(
m+|m|+1
\right)
\right]g_{m}
= 0. \label{sim2}
\end{eqnarray}
Using the change of variable $z=r^{2}/2 l_{B}^{2}$, the equation is rewritten in the form
\begin{eqnarray}
z \frac{d ^{2}g_{m}}{d z^{2}}+\left(|m|+1 - z \right)
\frac{d g_{m}}{dz}
\nonumber\\
+
\left[
\frac{E}{\hbar \omega_{c}}-\frac{1}{2}
\left(m+|m|+1
\right)
\right]g_{m}=0.
\end{eqnarray}

The confluent hypergeometric function $\Phi(\alpha, \gamma, z)$ obeys the same kind of equation
\begin{equation}
z \frac{d ^{2}\Phi}{d z^{2}}+\left( \gamma - z \right) \frac{ d \Phi}{dz}- \alpha
\Phi=0.
\end{equation}
There exist two linearly independent solutions to this differential equation
 $\Phi(\alpha, \gamma, z)$ and $z^{1 -\gamma} \Phi(\alpha-\gamma+1, 2-\gamma,z)$. The second solution diverges for $z=0$ and is therefore not physically acceptable. Furthermore, for $z \to \infty$

$$
\Phi(\alpha, \gamma, z) \sim \frac{\Gamma(\gamma)}{\Gamma(\alpha)} e^{z} z^{\alpha-\gamma}
$$
with $\Gamma(z)$ the gamma function.
Using the fact that the product   $r f_{m}(r)$ has to be square-integrable, we obtain that necessarily
$$\frac{1}{\Gamma(\alpha)}=0
$$
which leads to $\alpha=-l$ where $l$ is a positive integer. Then the confluent hypergeometric functions $\Phi(-l,|m|+1,z)$ are proportional to the generalized Laguerre polynomials
 $L_{l}^{|m|}(z)$ of degree $l$.  This results in the energy quantization
\begin{equation}
E=\hbar \omega_{c}\left(
l+\frac{1}{2}\left[|m|+m\right]+\frac{1}{2}
\right)=\hbar \omega_{c}\left(
n+\frac{1}{2}\right) \label{qu2}
\end{equation}
with $n=l+(|m|+m)/2$.
The corresponding wavefunctions are labelled by the quantum numbers
 $m$ and $l$
\begin{equation}
\Psi_{m,l}(r,\theta)=e^{im \theta} \, \left(\frac{r}{\sqrt{2}l_{B}}\right)^{|m|}
e^{-r^{2}/4 l_{B}^{2}} \,
L_{l}^{|m|}\left(\frac{r^{2}}{2 l_{B}^{2}}\right). \label{rot}
\end{equation}
Here, the moduli $|\Psi_{m,l}(r,\theta)|$ are rotation-invariant.
In contrast with the former basis, the present eigenstates are square-integrable in the plane and may thus correspond to physical states. The solutions (\ref{rot}) describe states for which there is an equally probable distribution of orbits whose centers lie on a circle \cite{Johnson}.

\subsection{Origin of the energy quantization?}

At this stage, let us compare the origin of the energy quantization in the two different derivations.
In the first case, the quantization originates from the vanishing of the wave function far from the origin in the $x$ direction.
According to the second derivation, two integral numbers contribute to the energy quantization: the first integer ($m$) results from the condition of single-valuedness of the wave function, while the second integer ($l$) stems from the condition of square-integrability.

From this comparison, the existence of a  physical interpretation for the energy quantization is not clear.
However, a clue for an unambiguous origin of this quantization seems to be provided by semi-classical arguments.
According to the classical equations of motion, a single electron confined to a plane and in the presence of a perpendicular magnetic field follows a circular orbit of radius $R_{c}=v/\omega_{c}$ where $v$ is a constant velocity.
The classical kinetic energy can thus be written as
\begin{equation}
E=\frac{1}{2} m^{\ast} v^{2}=\frac{1}{2} m^{\ast} (R_{c}\omega_{c})^{2}=\hbar \omega_{c} \frac{\Phi_{c}}{\Phi_{0}}
\end{equation}
with $\Phi_{c}=\pi R_{c}^{2} B$ the magnetic flux enclosed by the trajectory. Making the correspondence between this classical expression and the quantum expression [Eq. (\ref{qu1}) or (\ref{qu2})] of the kinetic energy,
we get in the semi-classical limit $n \gg 1$
\begin{equation}
\Phi_{c}=n \Phi_{0}.
\end{equation}
In this limit, the energy quantization can thus be interpreted as a consequence of the quantization of the flux enclosed by the cyclotron trajectory.
Furthermore, the flux enclosed by the trajectory can be seen  in semi-classical terms as the phase accumulated during a cycle.
We thus may wonder how  the energy quantization  is related to the phase of the wavefunction in fully quantum-mechanical terms.
In subsection \ref{wave}, we shall present a third derivation, which has the advantage to show explicitly
how the energy quantization arises {\it only} from a topological condition in the point of view of the electron's wave-particle duality.

\subsection{Gauge-invariant formulation}

In order to understand the difference between the two bases derived previously and the role of the gauge, it is useful to work in a gauge-invariant formalism. An important consequence of the presence of a magnetic field is that the hamiltonian $H_{0}$ is always complex whatever the gauge chosen for the vector potential. This implies the existence of a non-trivial phase for the wave function. Let us thus seek $\Psi$ in the form

$$ \Psi=f \exp(i \varphi)$$
with $f$ and $\varphi$ two real functions. Substituting this form in the equation (\ref{VP}) we get the system of coupled equations
\begin{eqnarray}
\bm{\nabla} \cdot \left[ f^{2}\left(\bm{\nabla} \varphi + \frac{2 \pi}{\Phi_{0}}{\bf A}\right)\right]=0 \label{system1a}
\\
-\Delta f+\left(\bm{\nabla} \varphi + \frac{2 \pi}{\Phi_{0}}{\bf A}\right)^{2}f
=\frac{2m^{\ast}E}{\hbar^{2}} f
\label{system1b}
\end{eqnarray}
where the functions $\varphi$ and $f$ have to be determined together with the energy $E$.
Note that in a hydrodynamical picture \cite{Madelung} of quantum mechanics (see also e.g. Ref. \onlinecite{jap}), $f^{2}$ plays the role of the density of a fictuous fluid and Eq. (\ref{system1a}) is nothing else than the equation of continuity for the current probability in the stationary case.

The explicit expression of the phase $\varphi$ of the wave function depends on the choice of a gauge.
Note that however the system (\ref{system1a})-(\ref{system1b}) is written in a gauge-invariant form. Indeed, if we consider the gauge transformation
  ${\bf A} \to {\bf A'}={\bf A}+\bm{\nabla} \Lambda$, the phase transforms as
\begin{equation}
\varphi \to \varphi'=\varphi-\frac{2\pi}{\Phi_{0}} \Lambda .\label{ph}
\end{equation}
The gauge is thus reabsorbed additively in the expression  (\ref{ph}) of the phase factor. Accordingly, the phase factor can be typically divided into two parts
$$\varphi=\varphi_{gd}+\varphi_{gi},$$
a part $\varphi_{gd}$ which is gauge dependent, and a part $\varphi_{gi}$ which is gauge independent.
Let us denote the combination of the phase gradient and the quantity $2\pi {\bf A}/\Phi_{0}$ by the gauge-invariant wavevector ${\bf K}$, i.e.
\begin{equation}
{\bf K}=\left(\bm{\nabla} \varphi + \frac{2 \pi}{\Phi_{0}}{\bf A}\right)
. \label{def}
\end{equation}
Eqs. (\ref{system1a})-(\ref{system1b}) are then rewritten as
\begin{eqnarray}
f \bm{\nabla} \cdot {\bf K} +2 \bm{\nabla} f  \cdot {\bf K}=0
\label{cont},
\\
-\Delta f+\left({\bf K}^{2}- \frac{2m^{\ast}E}{\hbar^{2}}\right)f
= 0 \label{eig}.
\end{eqnarray}
The coupling of the continuity equation (\ref{cont}) with the Eq. (\ref{eig}) was bypassed in the former derivations of the energy eigenvalues by invoking the choice of a gauge and symmetry considerations.
However, it is clear here that it is possible to get the energy quantization without specifying a gauge but rather by choosing the
 vectors
 ${\bf K}$.

 Indeed, for the translation-invariant states, we impose that the function $f$  depends only on the variable $x$.
If we impose in addition $\bm{\nabla} \cdot {\bf K}=0$, we obtain from Eq. (\ref{cont}) that ${\bf K} \parallel {\bf y}$. Then, the vectors ${\bf K}$ are fully determined from Eq. (\ref{def}) which yields
  $\bm{\nabla} \times {\bf K}= 2 \pi B  \hat{{\bf z}}/\Phi_{0}$
in the absence of phase singularities.
After integration, we obtain that
\begin{equation}
{\bf K}=
\left(p+ \frac{2 \pi x B}{\Phi_{0}}\right)\hat{{\bf y}},
\end{equation}
where $p$ is a constant of integration which is obviously gauge-independent.
Substituting this flow for ${\bf K}$ in Eq. (\ref{eig}), we get the gauge-independent Eq. (\ref{eq1}) yielding the energy spectrum.

For the rotation-invariant states, we impose that $f$ depends only on $r$ and
 $\bm{\nabla} \cdot {\bf K}=0$. Eq. (\ref{cont}) yields straightforwardly that
\begin{equation}
{\bf K} \parallel {\bf z} \times {\bf r}. \label{cond}
\end{equation}
In the presence of a phase singularity at the origin, we have
\begin{equation}
\bm{\nabla} \times {\bf K}= 2 \pi B  \hat{{\bf z}}/\Phi_{0}+2 \pi \, m \, \delta\left({\bf r}\right) \hat{{\bf z}} , \label{form}
\end{equation}
with $m$ a positive or negative integer ensuring the single-valuedness of the wave function $\Psi$.
The form (\ref{form}) together with the condition (\ref{cond}) determines also an admissible flow of ${\bf K}$ given by
\begin{equation}
{\bf K}=\left(\frac{m}{r^{2}}+
\frac{\pi B}{\Phi_{0}}\right)  \hat{{\bf z}} \times {\bf r}.
\end{equation}
In the symmetrical gauge, the phase associated with this flow is given by $\varphi=\varphi_{gi}=m \theta$, the gauge-dependent part $\varphi_{gd}$ being simply a constant.

Therefore, contrary to what could be inferred from the derivations \ref{Sec_trans} and \ref{Sec_rot}, it is not necessary to choose a gauge to write down the Schr\"{o}dinger's equation explicitly.
Furthermore, the real difference between the translation-invariant states and the rotation-invariant states is not the gauge because both kind of states can be obtained in any gauge. The real difference is in the choice of the flow of ${\bf K}$, which is intimately related to the symmetry of the probability density $f^{2}$. The set of quantum numbers labelling the eigenstates arises directly from this choice.
For instance, the vector ${\bf K}$ can be either translation invariant or rotation invariant.
As a result, there are many possibilities to choose other flows of ${\bf K}$ satisfying the system of equations (\ref{cont})-(\ref{eig}).
It is somehow remarkable that whatever this choice, the energy remains quantized as $(n+1/2)\hbar \omega_{c}$.

\subsection{Vortex eigenstates} \label{wave}

In absence of any potential, the electron
 follows classically a periodic cyclotron motion around an arbitrary center ${\bf R}$.
At the quantum level, the translation invariance of the hamiltonian $H_{0}$ reflects precisely the degeneracy of the kinetic energy with respect to the position of the center, while its rotation invariance corresponds to the cyclotron motion.
Apparently, the former eigenstates (\ref{trans}) or (\ref{rot}) do not reflect the symmetry of the cyclotron motion around an arbitrary point ${\bf R}$ so that the consideration of the classical limit is rather tricky.
Since they do not correspond to the classical picture of the motion, it is difficult to appreciate the wave-particle duality.

Let us thus try to construct a basis of eigenstates which has the right properties.
For this purpose, we impose that the probability density $f^{2}$ has the same symmetry as the cyclotron motion, i.e. is a function of $|{\bf r}-{\bf R}|$ only. As previously, we take the freedom of having $\bm{\nabla} \cdot {\bf K}=0$.
Then,  we get from Eq. (\ref{cont}) that
\begin{equation}
{\bf K} \parallel {\bf z} \times ({\bf r}-{\bf R}).
\end{equation}
It is straightforward to see that
the flow
\begin{equation}
{\bf K}= \left(\frac{m}{|{\bf r}-{\bf R}|^{2}} +\frac{\pi B}{\Phi_{0}} \right)  \hat{{\bf z}} \times ({\bf r}-{\bf R}) \label{flow}
\end{equation}
is admissible from the point of view of the coupling of Eqs. (\ref{cont})-(\ref{eig}). Here, we have deliberately located the phase singularity at the arbitrary position ${\bf R}$ (rather than at the origin as in \ref{Sec_rot}).

Substituting the form (\ref{flow}) in the Eq. (\ref{eig}) we obtain
a differential equation for $f$ which is similar to Eq. (\ref{sim}) after introducing the  new cylindrical coordinates defined by
$\rho=|{\bf r}-{\bf R}|$ and
$\Theta=\arg\left({\bf r} -{\bf R}\right)$.
The function
\begin{equation}
f_{m, {\bf R}}({\bf r})= \rho^{m} \, e^{-\rho^{2}/4 l_{B}^{2}}
\end{equation}
with $m \geq 0$ is a simple solution of this differential equation yielding the energy quantization
\begin{equation}
E_{m}=\hbar \omega_{c}\left(
m+\frac{1}{2}\right). \label{eigenv}
\end{equation}
At this stage, it is not needed to entirely solve the differential equation (\ref{sim})
as done in \ref{Sec_rot} because we have already determined by this way enough states. For the same reason, we have restricted ourself to positive values of the integer $m$.

Then, considering Eq. (\ref{flow}) and taking the symmetrical gauge ${\bf A}={\bf B} \times {\bf r}/2$,
we deduce that
the phase-gradient  takes the form
\begin{equation}
{\bm \nabla} \varphi= m \hat{{\bf z}} \times \frac{{\bf r}-{\bf  R}}{|{\bf r}-{\bf  R} |^{2}} -\frac{\hat{{\bf z}} \times {\bf R}}{2 l_{B}^{2}}.
\end{equation}
We obtain that the normalized wave functions labelled by $m$ and ${\bf R}=(X,Y)$ are fully expressed  in the symmetrical gauge as
\begin{eqnarray}
\Psi_{m,{\bf  R}}({\bf r})&=&\frac{C_{m}}{l_{B}} \left|\frac{{\bf r}-{\bf R}}{\sqrt{2}l_{B}} \right|^{m} \!\! e^{i m \, \mathrm{arg}({\bf r}-{\bf R})} \, e^{-\frac{({\bf r}-{\bf R})^{2}-2 i \hat{{\bf z}} \cdot ( {\bf r} \times {\bf R})}{4 l_{B}^{2}}} \nonumber
\\
&=&\frac{C_{m}}{l_{B}} \left(\frac{z-Z}{\sqrt{2}l_{B}}\right)^{m}  \,
 e^{-\frac{|z|^{2} +|Z|^{2}-2 Z z^{\ast}}{4 l_{B}^{2}}} \label{vortex}
\end{eqnarray}
with $C_{m}=1/\sqrt{2\pi m!}$, $z=x+iy$, and $Z=X+i Y$.
For practical convenience, we use henceforth the Dirac notation
$$\Psi_{m,{\bf R}}({\bf r})=\langle {\bf r} | m, {\bf R} \rangle .$$

We now argue that the set of eigenstates  $| m, {\bf R} \rangle$ with $m$ restricted to positive values only ($m \geq 0$) forms a semi-orthogonal basis.
The overlap between two states is calculated in Appendix A, with the result:
\begin{eqnarray}
 \langle m, {\bf R} | m', {\bf R}' \rangle=\delta_{m \,m'} \,
e^{-\frac{
({\bf R}-{\bf R}')^{2} -2 i \hat{{\bf z}} \cdot ({\bf R} \times {\bf R}')}{4 l_{B}^{2}} }. \label{overlap}
\end{eqnarray}
According to (\ref{overlap}), the states $| m, {\bf R} \rangle$ are orthogonal  with respect  to the quantum number $m$, but not with respect to the vortex positions ${\bf R}$. However, the overlap is almost zero when the distance between two vortex positions exceeds a few magnetic lengths.
Let us prove that the set of vortex states (\ref{vortex}) is complete. We have

\begin{widetext}
\begin{eqnarray}
\int d^{2}{\bf R} \sum_{m=0}^{+\infty} \langle {\bf r} | m, {\bf R}  \rangle \langle m, {\bf R} |{\bf r}' \rangle=
\int \!\!\! \frac{ d^{2}{\bf R}}{2 \pi l_{B}^{2}} \sum_{m=0}^{+\infty}
\frac{1}{m!}
\left(\frac{z'^{\ast}-Z^{\ast}}{\sqrt{2}l_{B}}
\right)^{m} \left(\frac{z-Z}{\sqrt{2} l_{B}}
\right)^{m} \,
 e^{-\frac{|z|^{2} +|z'|^{2}+2|Z|^{2}-2 Z z^{\ast}-2Z^{\ast} z'}{4 l_{B}^{2}}}.
\end{eqnarray}
Recognizing the development into series  of the exponential function, we perform first the sum over $m$ to get
\begin{eqnarray}
\int d^{2}{\bf R} \sum_{m=0}^{+\infty}  \langle {\bf r} | m, {\bf R}  \rangle \langle m, {\bf R} |{\bf r}' \rangle =
e^{-\frac{|z|^{2}+|z'|^{2}-2z'^{\ast} z}{4 l_{B}^{2}}}
\int \!\! \frac{d^{2}{\bf R}}{2 \pi l_{B}^{2}}  \, e^{-i \mathrm{Im} \, \frac{Z^{\ast}(z-z')}{l_{B}^{2}}}.
\end{eqnarray}
\end{widetext}
The remaining integrations over the vortex coordinates ${\bf R}$ yield straightforwardly
\begin{eqnarray}
\int d^{2} {\bf R} \sum_{m=0}^{+\infty}  \langle {\bf r} | m, {\bf R}  \rangle \langle m, {\bf R} |{\bf r}' \rangle = 2 \pi l_{B}^{2} \,\delta({\bf r}-{\bf r}'). \label{clo1}
\end{eqnarray}
Using that $\delta({\bf r}-{\bf r}')=\langle { \bf r} | {\bf r}' \rangle$, inserting from the left $|{\bf r} \rangle$ and from the right $\langle {\bf r}'|$, integrating over ${\bf r}$ and ${\bf r}'$, and using the completeness relation for the states $|{\bf r} \rangle$, we get from this expression (\ref{clo1}) the completeness relation
\begin{eqnarray}
\int \frac{d^{2}{\bf R}}{2 \pi l_{B}^{2}} \sum_{m=0}^{+\infty} \left| m,{\bf R}\rangle \langle  m,{\bf R} \right| = 1. \label{closure}
\end{eqnarray}
We see that the set of states is complete if we associate the area $2 \pi l_{B}^{2}$ with the vortex coordinates $X$ and $Y$.
We deduce that the degeneracy of the energy levels is equal to $(2 \pi l_{B}^{2})^{-1}$ per unit area (i.e. for a finite system it corresponds to $\Phi/\Phi_{0}$ where $\Phi$ is the total magnetic flux in the system), in agreement with the standard derivations making use of orthogonal sets of eigenstates.

We have therefore presented a third way of derivation of the eigenvalues and eigenstates for the problem of a charged particle in an uniform magnetic field.
We find that the energy eigenvalues (\ref{eigenv})
are
 degenerate with respect to a continuous quantum number, namely the positions ${\bf  R}$, which here correspond to the locations of a phase singularity. In the limit of infinite magnetic field only, the vortex location ${\bf R}$ corresponds to the classical guiding center coordinates.
It is worth noting that, contrary to previous standard derivations,  we
  did not resort to any condition of square-integrability
 to obtain the quantization of the energy levels.
Here, the quantization of the kinetic energy stems {\it only} from the quantization of the angular momentum (defined with respect to the position ${\bf R}$).
In terms of the wave-particle duality, the mechanism at hand is clearly the interference of the electronic wavefunction with itself due to the completion of a circular orbit.
 The eigenstates correspond to vortices located at the positions ${\bf R}$. Their non-orthogonality with respect to ${\bf R}$ reflects  the quantum indeterminacy ($2 \pi l_{B}^{2}$ according to the completeness relation) in the positions of the vortices. Eq. (\ref{closure}) constitutes an essential relation that allows to expand any wave function into a superposition of the vortex states $|m, {\bf R} \rangle$.

Because one of the quantum numbers ($m$) results only from the single-valuedness of the wave function, i.e. stems from a topological constraint, it is sensitive to local perturbations only on the scale of the magnetic length around the vortex positions ${\bf R}$.
This property helps for example to capture the interplay of a confining potential and the Landau levels in a straightforward manner.
Within the vortex states basis, it is clear
that only the states whose vortex positions lie near the edges will be affected by the confinement.  As a consequence, a  deviation of the spectrum from the Landau levels may be only significant near the edges.
Similar conclusions concerning the spectrum of finite systems can be reached  after thorough calculations for particular models as the Teller model (confinement by hard walls) or the  model of elastic confinement (for a discussion, see e.g. Ref. \onlinecite{Hadju}).
 On the other hand, the bulk states are not affected by the confining potential, which justifies the common disregard of the influence of edges when computing thermodynamical quantities of macroscopic systems.
One important advantage of the present basis is therefore to allow a separate quantum treatment of the edges and of the bulk.
Despite the lack of orthogonality, we shall show in the next Section that the basis of vortex states is also very convenient for the treatment of any random potential that varies smoothly on the scale of the magnetic length  in high magnetic fields. 

Finally, we note that the set of states (\ref{vortex}) has been derived previously \cite{Malkin1969,Materdey2003} in the literature within different approaches and contexts, namely the coherent states formalism developed by Glauber \cite{Glauber1963} (for a review on the theory of coherent states see e.g. Ref. \onlinecite{Zhang1990}). Indeed, the states (\ref{vortex}) can be identified as coherent states of a charged particle in a magnetic field and they naturally obey the algebra \cite{Glauber1963,Zhang1990} expected for these states (the algebra is characterized by the identities (\ref{overlap}) and (\ref{clo1})).
A peculiarity here is that the vortex states (\ref{vortex}) are both coherent states (this property comes out through the degeneracy quantum number ${\bf R}$ - the vortex location) and eigenstates of the hamiltonian (the eigenvalues are associated with the quantum number $m$). \cite{Note}
The topological nature of the energy quantum number $m$, that is fundamental to capture the quantum transport properties in finite magnetic fields as will be developed in the present paper, has seemingly not been recognized in these approaches. \cite{Malkin1969,Materdey2003}
The set of states  (\ref{vortex}) has already been used for the calculation of  the partition function \cite{Apenko1983} and the Wigner function. \cite{Materdey2003}
We shall present a systematic expansion using this set in the framework of the
real-time Green functions to study {\it both} thermodynamic and quantum transport properties in high magnetic fields.

\vspace*{0.5cm}

\section{Effect of a smooth potential on  Landau levels in high magnetic fields}

\subsection{Matrix elements of the potential}

In this Section, we consider that the electron feels, in addition to the perpendicular magnetic field, the presence of a potential $V(x,y)$.
The corresponding Hamiltonian is therefore
\begin{eqnarray}
H=H_{0}+ V(x,y), \label{Hamil}
\end{eqnarray}
where $H_{0}$ is the kinetic part introduced in the former Section.
The matrix elements
of the potential $V$ expressed in the semi-orthogonal basis $|m,{\bf R} \rangle$ of eigenstates of $H_{0}$ are given by

\begin{eqnarray}
V_{m,m'}({\bf R},{\bf R'})
&=& \int
 d^{2}{\bf r} \, V({\bf r}) \,
\Psi_{m,{\bf R}}^{\ast}({\bf r}) \,\Psi_{m',{\bf R}'}({\bf r}).
\end{eqnarray}
We need to arrange the expression of these matrix elements by making some manipulations. For this purpose,
we shall consider that $V({\bf R})$ can be expanded in a Taylor series
\begin{equation}
V({\bf r})= \sum_{j=0}^{+\infty}  \frac{1}{j!}
\left.
\left[{\bf r} \cdot {\bm \nabla}_{{\bf u}} \right]^{j} V({\bf u}) \right|_{{\bf u}={\bf 0}}.
\end{equation}

Introducing the relative coordinates ${\bf d}= ({\bf R}'-{\bf R})/2$ and the center of mass coordinates ${\bf c}=({\bf R}'+{\bf R})/2$, we first make the change of variable ${\bf r}'=({\bf r}-{\bf c}-i {\bf d} \times \hat{{\bf z}})/\sqrt{2}l_{B}$
\begin{widetext}
\begin{eqnarray}
V_{m,m'} ({\bf R},{\bf R}')
&=&2  C_{m} C_{m'} \langle {\bf R}| {\bf R}' \rangle \int \!
\! d x' \!\! \int\!\! d y'\,
\left[x'-i y'
\right]^{m} \left[x'+i y' \right]^{m'}
 V\left( \sqrt{2} l_{B} {\bf r}'+{\bf c}+i {\bf d} \times \hat{{\bf z}}\right) \, e^{-(x'^{2}+y'^{2})} \label{course1}
\end{eqnarray}
where  $\langle {\bf R}| {\bf R}' \rangle =\langle m, {\bf R}| m, {\bf R}' \rangle$, and $x'$ (respectively $y'$) lies in the complex plane  on the line defined by $z=-i d_{y}/\sqrt{2} l_{B}$ ($z= i d_{x}/\sqrt{2} l_{B}$). The potential term is then expanded in Taylor series around the complex point ${\bf c}+i {\bf d} \times \hat{{\bf z}}$
\begin{equation}
V\left( \sqrt{2} l_{B} {\bf r}'+{\bf c}+i {\bf d} \times \hat{{\bf z}}\right)= \sum_{j=0}^{+\infty}  \frac{(\sqrt{2}l_{B})^{j}}{j!}
\left.
\left[{\bf r}' \cdot {\bm \nabla}_{{\bf u}} \right]^{j} V({\bf u}) \right|_{{\bf u}={\bf c}+i{\bf d} \times \hat{{\bf z}}}. \label{ser}
\end{equation}
Now, if the length scale $\lambda$ characterizing the range over which the potential $V$ varies significantly is typically bigger than the magnetic length $l_{B}$ (this occurs for any nonpathological potentials in sufficiently high magnetic fields), the series (\ref{ser}) converges for all finite $x'$ and $y'$ and represents a function which is analytic throughout the complex plane with respect to these two complex variables. Using this analycity property, the contours of integration can be deformed to the real axes.
Then,
noting that
\begin{equation}
2 {\bf r}' \cdot {\bm \nabla}_{{\bf u}}
=(x'-iy')\left(\partial_{\eta_{x}}+i \partial_{\eta_{y}} \right) +(x'+iy')\left(\partial_{\eta_{x}}-i \partial_{\eta_{y}} \right),
\end{equation}
with ${\bf u}=(\eta_{x},\eta_{y})$, and using the binomial theorem, we get after integration that the matrix elements of the potential can be written as
\begin{eqnarray}
V_{m,m'}({\bf R},{\bf R}') = \sum_{j=0}^{+ \infty}  \sum_{k=0}^{j} \frac{\langle {\bf R} | {\bf R}' \rangle }{k!(j-k)!} \left(\frac{l_{B}}{\sqrt{2}}\right)^{k}  \frac{(m+k)!}{\sqrt{m!m'!}}
\, \delta_{m+k,m'+j-k} \left.
(\partial_{\eta_{x}}+i\partial_{\eta_{y}})^{k}(\partial_{\eta_{x}}-i\partial_{\eta_{y}})^{j-k}
 V(\eta_{x},\eta_{y}) \right|_{(\eta_{x},\eta_{y})={\bf c}+i {\bf d} \times \hat{{\bf z}}} . \label{aroundc}
\end{eqnarray}
 Then, using the fact that ${\bf c}+i {\bf d} \times \hat{{\bf z}}={\bf R}+{\bf d}+i {\bf d} \times \hat{{\bf z}}$ and applying once again the Taylor expansion to have the potential and its higher order derivatives taken at the point ${\bf R}$ only,
 we reorganize the different terms appearing in the series (\ref{aroundc}) with respect to the order $q$ of the derivatives of the potential $V$ as

\begin{eqnarray}
V_{m,m'}({\bf R},{\bf R}') &=&   \langle {\bf R} | {\bf R}' \rangle \sum_{q=0}^{+ \infty} V_{m,m'}^{(q)}({\bf R},{\bf R}'). \label{series}
\end{eqnarray}
The leading contribution to the series is
\begin{equation}
V^{(0)}_{m,m'}({\bf R},{\bf R}')=V\left({\bf R}\right) \delta_{m,m'}.
\end{equation}
For the subsequent calculations (Section V), we shall need also the expression for $q=1$ which can be written as
\begin{eqnarray}
V_{m,m'}^{(1)}({\bf R},{\bf R}') &=&
\left\{
\frac{({\bf R}'-{\bf R})}{2} \cdot {\bm \nabla} V ({\bf R})-
i \hat{z} \cdot \left[\frac{({\bf R}'-{\bf R})}{2} \times {\bm \nabla} V ({\bf R})
\right]
 \right\}
\,
\delta_{m,m'}+
\nonumber  \\
&&
+
l_{B} \sqrt{\frac{m'}{2}}\left\{ \partial_{x} V ({\bf R})+i\partial_{y} V ({\bf R})\right\}\, \delta_{m+1,m'}
+
l_{B} \sqrt{\frac{m}{2}}\left\{ \partial_{x} V({\bf R})-i\partial_{y} V({\bf R})\right\}\, \delta_{m,m'+1}
. \label{mV}
\end{eqnarray}
\end{widetext}

We are now in a position to analyze the peculiarities of a smooth potential in the high magnetic field regime.
Due to the exponential cutoff factor $\langle {\bf R} | {\bf R}'\rangle$, the matrix elements $V_{m,m'}({\bf R},{\bf R}')$ are non negligible only when ${\bf R}'$ is  located at a distance from ${\bf R}$ which is typically smaller than $\sqrt{2} l_{B}$, i.e. the modulous $|{\bf d}|$ is at best of the order of $l_{B}$.
We deduce that
\begin{equation}
\left| V_{m,m'}^{(q)}({\bf R},{\bf R}') \right| \sim \left(\frac{l_{B}}{\lambda} \right)^{q} V({\bf R})
\end{equation}
where $\lambda$ is the length scale characterizing the spatial variations of $V$ around the point ${\bf R}$. Therefore,  when the potential $V$ varies smoothly on the scale of the magnetic length $l_{B}$, $\lambda \gg l_{B}$, we have the inequalities
\begin{eqnarray}
\left| V_{m,m'}^{(0)}({\bf R},{\bf R}') \right| \gg \left| V_{m,m'}^{(1)}({\bf R},{\bf R}') \right| \gg \left| V_{m,m'}^{(2)}({\bf R},{\bf R}') \right| ... \nonumber
\end{eqnarray}
This shows that we have ordered in the series (\ref{series}) the different contributions to $V_{m,m'}({\bf R},{\bf R}')$ by their magnitude of relevance (characterized by $q$).
In the next subsection, we exploit this fact to build a systematic gradient expansion  within the formalism of Green functions.


Finally, it is worth noting that the vortex states (\ref{vortex}) are labeled by the continuous index ${\bf R}$, and therefore the set $|m,{\bf R} \rangle$ is over-complete in the Hilbert space that has a countable basis. Nevertheless, according to the completeness relation (\ref{closure}) it is possible to expand any arbitrary state or operators in the vortex states representation. This is precisely what we have done here for the potential. Still, it is in principle necessary to be careful at this level because the vortex states are obviously not linearly independent of one another, and thus the expansion of $V$ in terms of these vortex states may be not unique.
However, clearly, the present expansion (\ref{series}) which is only valid for a smooth potential in the high magnetic field regime (more precisely the condition is related to the criterion of convergence of the series (\ref{ser})) is unambiguous: there is a unique correspondence between the matrix elements and the operator $V$.
In the framework of the coherent states theory \cite{Glauber1963,Zhang1990} the unicity of the expansion is provided by the constraint that the expansion coefficents depends analytically upon the continuous (complex) quantum number labelling the states.
Here, we have used explicitly the same condition in the course of the derivation (between Eq. (\ref{course1}) and (\ref{aroundc})) of the matrix elements of the potential. Namely, the criterion of convergence of the Taylor expansion of the potential has implied the analycity of the function $V$, what renders the expansion unique.

\subsection{Green functions formalism}

To investigate thermodynamic and transport properties of the two-dimensional electron gas in a smooth potential $V$ and in high magnetic fields, we shall use the formalism of real-time  Green functions. Retarded, advanced and Keldysh Green functions are respectively defined as (see e.g. Refs. \onlinecite{Mahan} and \onlinecite{Rammer})
\begin{eqnarray}
G^{R}(x_{1},x_{2}) & = & - i \theta(t_{1}-t_{2}) \, \left \langle \left\{
\Psi(x_{1}),\Psi^{\dag}(x_{2})
\right\} \right \rangle \\
G^{A}(x_{1},x_{2}) & = & i \theta(t_{2}-t_{1}) \, \left \langle \left\{
\Psi(x_{1}),\Psi^{\dag}(x_{2})
\right\} \right \rangle \\
G^{K}(x_{1},x_{2}) & = & -i \left \langle \left[
\Psi(x_{1}),\Psi^{\dag}(x_{2})
\right] \right \rangle
\end{eqnarray}
where $[\,,\,]$ means the commutator, $\{\,,\,\}$ the anticommutator, and $\theta$ is the Heaviside function.
The Green functions relate the field operator $\Psi(x)$ of the particle at one point $x_{1}=({\bf r}_{1},t_{1})$ in space-time to the conjugate field operator $\Psi^{\dag}(x_{2})$ at another point $x_{2}=({\bf r}_{2},t_{2})$. The field operators $\Psi(x_{1})$ and $\Psi^{\dag}(x_{2})$ are expressed in terms of the eigenfunctions $\Psi_{\nu}({\bf r})$ and eigenvalues $\varepsilon_{\nu}$ as
\begin{eqnarray}
\Psi(x_{1})&=&\sum_{\nu} c_{\nu} \Psi_{\nu}({\bf r}_{1}) \, e^{-i \varepsilon_{\nu}t_{1}}\\
\Psi^{\dag}(x_{2})&=&\sum_{\nu} c_{\nu}^{\dag} \Psi_{\nu}^{\ast}({\bf r}_{2}) \, e^{i \varepsilon_{\nu}t_{2}}
\end{eqnarray}
where $c_{\nu}^{\dag}$ and $c_{\nu}$ are respectively the creation and destruction operators.

In the absence of the potential $V$, the Green functions can be written explicitly. We denote them by $G_{0}$. Due to its relevance for the high magnetic field expansion theory, we shall use the basis of eigenstates $|\nu \rangle = |m, {\bf R} \rangle$
derived in Section II.E, which is complete with the prescription (see the closure relation (\ref{closure}))
\begin{equation}
\sum_{\nu} =\sum_{m=0}^{+\infty} \int \!\! \frac{d^{2} {\bf R}}{2 \pi l_{B}^{2}}.
\end{equation}
The different Green functions $G_{0}$ are thus given by
\begin{eqnarray}
G^{R}_{0}(x_{1},x_{2}) &=&
-i \theta(t) \sum_{\nu}
\Psi_{\nu}({\bf r}_{1}) \Psi_{\nu}^{\ast}({\bf r}_{2})
e^{-i \varepsilon_{\nu}t},
\\
G^{A}_{0}(x_{1},x_{2}) &=&
i \theta(-t) \sum_{\nu}
\Psi_{\nu}({\bf r}_{1}) \Psi_{\nu}^{\ast}({\bf r}_{2})
e^{-i \varepsilon_{\nu}t}
,
\\
\nonumber
G^{K}_{0}(x_{1},x_{2}) &=&  i \sum_{\nu}
(2n_{F}(\xi_{\nu})-1) \Psi_{\nu}({\bf r}_{1}) \Psi_{\nu}^{\ast}({\bf r}_{2})
e^{-i \varepsilon_{\nu}t}\\
\end{eqnarray}
where $t=t_{1}-t_{2}$ and
\begin{equation}
n_{F}(\xi_{\nu})=\langle c^{\dag}_{\nu} c_{\nu} \rangle=\frac{1}{1+\exp(\xi_{\nu}/T)}
\end{equation}
is the Fermi-Dirac distribution function with $\xi_{\nu}=\varepsilon_{\nu}-\mu$ and  $\varepsilon_{\nu}=(m+1/2)\hbar \omega_{c}$.

Furthermore, instead of working in the representation position $\left | {\bf r} \right. \rangle $, we shall henceforth work in the representation $\left | \nu  \right.\rangle$  where  the Green functions take a simpler form due to the semi-orthogonality property of the basis $\nu$

\begin{eqnarray}
G^{R}_{0}(\nu_{1},t_{1};\nu_{2},t_{2}) &=&
-i \theta(t) \, \langle \nu_{1} | \nu_{2} \rangle
e^{-i \varepsilon_{\nu_{1}}t} ,
\\
G^{A}_{0}(\nu_{1},t_{1};\nu_{2},t_{2}) &=&
i \theta(-t)  \,
 \langle \nu_{1} | \nu_{2} \rangle
e^{-i \varepsilon_{\nu_{1}}t} ,
\\
G^{K}_{0}(\nu_{1},t_{1};\nu_{2},t_{2}) &=&  i
\left(2n_{F}(\xi_{\nu_{1}})-1\right) \,
 \langle \nu_{1} | \nu_{2} \rangle
e^{-i \varepsilon_{\nu_{1}}t}
.
\end{eqnarray}
The Green functions are diagonal with respect to the energy quantum number $m$
but not with respect to the degeneracy quantum number ${\bf R}$.
After Fourier transformation with respect to the time difference $t$, the Green functions are written in the energy ($\omega$) representation as
\begin{eqnarray}
G^{R}_{0}(\nu_{1};\nu_{2}) &=&
\frac{ \delta_{m_{1},m_{2}} \, \langle {\bf R}_{1}|{\bf R}_{2}\rangle}{\omega-\xi_{m_{1}}  + i \delta}  ,
\\
G^{A}_{0}(\nu_{1};\nu_{2}) &=&
\frac{\delta_{m_{1},m_{2}} \, \langle {\bf R}_{1}|{\bf R}_{2}\rangle}{\omega-\xi_{m_{1}} -i \delta}
 ,
\\
\nonumber
G^{K}_{0}(\nu_{1} ;\nu_{2}) &=&  -2 i \pi  \tanh\left(\frac{\omega}{2T}\right)
\, \langle \nu_{1}|\nu_{2}\rangle
\delta(\omega-\xi_{m_{1}}).\\
\end{eqnarray}
For many calculations with the basis $|\nu \rangle $, it is useful to note that within each eigensubspace, we have the following identity
\begin{equation}
\int \frac{d^{2} {\bf R}}{2 \pi l_{B}^{2}}\langle m, {\bf R}'   |m,{\bf R}\rangle \langle m,{\bf R}| m, {\bf R}'' \rangle=\langle m,{\bf R}'|m, {\bf R}'' \rangle 
\end{equation}
which can be interpreted as a closure relation that holds in the fixed $m$ eigensubspace
\begin{equation}
\int \frac{d^{2} {\bf R}}{2 \pi l_{B}^{2}} |m,{\bf R}\rangle \langle m,{\bf R}| =1_{m}. \label{clom}
\end{equation}

\subsection{High magnetic fields gradient expansion: retarded and advanced Green functions}

The retarded and advanced Green functions in the presence of the perturbation $V$ are obtained from the Dyson equation, which takes the form in the energy ($\omega$) representation (for more details see Appendix B.1)

\begin{equation}
(\omega-\xi_{m_{1}}\pm i \delta) \, G^{R,A}_{\nu_{1},\nu_{2}}(\omega)= \langle \nu_{1} |\nu_{2} \rangle+\sum_{\nu_{3}} V_{\nu_{1},\nu_{3}}G^{R,A}_{\nu_{3},\nu_{2}}(\omega). \label{DysonRA}
\end{equation}
Note that the quantum-kinetic equation obeyed by the Keldysh Green function $G^{K}$ in the representation $|\nu \rangle$ is derived  in the Appendix B.1. The function $G^{K}$ important for quantum transport properties is determined in the Appendix B.2.

Here $V$ includes both the scalar electrostatic potential and the impurity potential. The explicit form of $V$ does not matter at this step. We only assume that $V$ is a smooth potential in high magnetic fields.
As mentioned above, the potential term $V_{\nu_{1} \nu_{2}}$ is thus principally described by the first terms in the series expansion (\ref{series}).
The retarded and advanced Green functions are also sought as a series expansion  of terms of the order of $(l_{B}/\lambda)^{q}$

\begin{equation}
G_{\nu_{1} \nu_{2}}=\sum_{q=0}^{+ \infty} G^{(q)}_{\nu_{1} \nu_{2}}.
\end{equation}

Inserting the leading contribution to $V_{\nu_{1} \nu_{2}}$ in the Dyson equation (\ref{DysonRA}), the zero order Green functions $G^{(0)}$ (in the presence of the potential) are given by
\begin{widetext}
\begin{equation}
(\omega-\xi_{m_{1}}\pm i \delta) \, G^{(0)\, R,A}_{\nu_{1},\nu_{2}}(\omega)= \langle \nu_{1} |\nu_{2} \rangle+\sum_{\nu_{3}} V^{(0)}_{\nu_{1},\nu_{3}}
\langle {\bf R}_{1} | {\bf R}_{3} \rangle
G^{(0)\, R,A}_{\nu_{3},\nu_{2}}(\omega)
\end{equation}
with $V^{(0)}_{\nu_{1},\nu_{3}}=V({\bf R}_{1}) \, \delta_{m_{1},m_{3}}$.
The summation over $\nu_{3}$ is readily performed using the relation (\ref{clom}) so that we get
\begin{equation}
G^{(0)\, R,A}_{\nu_{1} \nu_{2}}(\omega)= \frac{\delta_{m_{1},m_{2}} \, \langle {\bf R}_{1}|{\bf R}_{2}\rangle}{\omega-\xi_{m_{1}}-V\left({\bf R}_{1}\right) \pm i \delta} . \label{exp0}
\end{equation}
We note that in the denominator of the Green functions (\ref{exp0}), the potential can be taken either at the point ${\bf R}_{1}$ or at the point ${\bf R}_{2}$ since, due to the presence of the overlap $\langle {\bf R}_{1}| {\bf R}_{2} \rangle$ in the numerator, $V({\bf R}_{1})$ differs from $V({\bf R}_{2})$ at best by a correcting term of the order of $V l_{B}/\lambda $ which can be disregarded in the zero order expression of the Green function.
The absence at leading order  of transitions between the quantum numbers $m$ is a clear manifestation of their purely topological (and local) nature: they are immune to small perturbations. The leading effect of the perturbation $V$ is that the vortex state acquires a local potential energy $V({\bf R})$ which  is responsible for a lifting of the Landau levels degeneracy with respect to the quantum number ${\bf R}$.

The next order Green functions $G^{(1)}$ in the presence of the potential
are given by
\begin{equation}
\left(\omega-\xi_{m_{1}}-V({\bf R}_{1}) \pm i \delta\right)
 G^{(1) \, R,A}_{\nu_{1} \nu_{2}}  (\omega)
 = \sum_{\nu_{3}}
V^{(1)}_{\nu_{1} \nu_{3}}
\langle {\bf R}_{1}| {\bf R}_{3} \rangle
 G^{(0) \, R, A}_{\nu_{3} \nu_{2}}  (\omega). \label{Dyorder1}
\end{equation}
Inserting the expression (\ref{exp0}) for $G^{(0)}$ with the potential $V$ in the denominator taken at the point ${\bf R}_{2}$, and performing the summation over ${\bf R}_{3}$ in the right-hand side of (\ref{Dyorder1}) by considering that
\begin{equation}
\int \frac{d^{2} {\bf R}_{3}}{2 \pi l_{B}^{2}} {\bf R}_{3} \, \langle {\bf R}_{1}|{\bf R}_{3} \rangle  \langle {\bf R}_{3}|{\bf R}_{2} \rangle =  \frac{1}{2}\left[
{\bf R}_{2}+{\bf R}_{1}- i \hat{{\bf z}} \times \left( {\bf R}_{2}-{\bf R}_{1}
\right)
\right]
\langle {\bf R}_{1}|{\bf R}_{2} \rangle
,
\end{equation}
we obtain
\begin{eqnarray}
G^{(1)\, R,A}_{\nu_{1} \nu_{2}}(\omega)=\frac{V^{(1)}_{m_{1},m_{2}}({\bf R}_{1},{\bf R}_{2}) \, \langle {\bf R}_{1}| {\bf R}_{2} \rangle }{(\omega-\xi_{m_{1}}-V({\bf R}_{1}) \pm i\delta)(\omega-\xi_{m_{2}}-V({\bf R}_{2}) \pm i \delta)}
.
\end{eqnarray}
\end{widetext}

\section{Density of states and chemical potential}

In this Section, we consider an equilibrium situation where
the potential energy $V({\bf R})$ consists of an impurity potential only.
Our goal is to investigate thermodynamical features such as the dependence of the chemical potential on the magnetic field under the regime of a smooth potential in high magnetc fields. For this purpose, we first derive the energy dependence of the density of states.

\subsection{Density of states}

\begin{figure}
\includegraphics[width=\columnwidth]{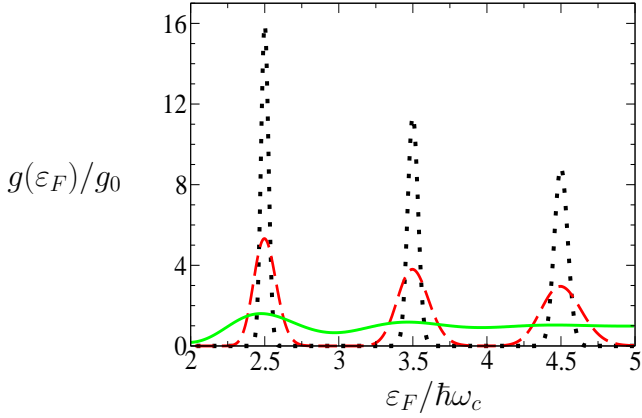}
\caption{ (color online)
Density of states at $\varepsilon_{F}$ as a function of the inverse magnetic field for various disorder broadening parameters $\Gamma$. The solid curve is obtained for $\Gamma= 0.1 \, \varepsilon_{F}$, the dashed curve for $\Gamma= 0.03 \, \varepsilon_{F}$, and the dotted curve for $\Gamma=0.01 \, \varepsilon_{F}$.  }
\label{dosc}
\end{figure}

The density of states $g(\varepsilon)$ at the energy $\varepsilon$ is obtained from
\begin{equation}
g(\varepsilon)=-\frac{1}{\pi} \mathrm{Im}   \sum_{\nu} G^{R}_{\nu,\nu}(\varepsilon)
\end{equation}
where $\varepsilon=\omega+\mu$.
Inserting the zero order Green function, the total density of states is straightforwardly given at leading order of the expansion theory by
\begin{equation}
g(\varepsilon)=\sum_{m=0}^{+\infty} \int \!\! \frac{d^{2} {\bf R}}{2 \pi l_{B}^{2}} \, \, \delta\left(\varepsilon - (m+1/2)\hbar \omega_{c}-V({\bf R})\right).
\end{equation}
Note that the next correction term to the density of states that we do not give here (although it can be  derived in a systematic way rather easily from our formalism) is provided by the second-order Green function $G^{(2)}$.
Introducing
\begin{equation}
\rho(V)=\int \!\! \frac{d^{2} {\bf R}}{S} \, \delta \left( V-V({\bf R}) \right)
\end{equation}
where $S$ is the total area of the system,
the density of states takes the form
\begin{equation}
g(\varepsilon)=\frac{S}{2 \pi l_{B}^{2}}\sum_{m=0}^{+\infty} \int \!\! d V \rho(V)\, \delta\left(\varepsilon - (m+1/2)\hbar \omega_{c}-V\right).
\end{equation}
The distribution of the amplitudes $V$ depends on the model considered.
 If we assume a gaussian distribution to account for the effect of a disordered potential present in the bulk
\begin{equation}
\rho(V)= \frac{1}{\sqrt{2 \pi} \Gamma} e^{-V^{2}/2 \Gamma^{2}},
\end{equation}
the density of states corresponds obviously to a sum of gaussian peaks
\begin{equation}
g(\varepsilon)=\frac{S}{2 \pi l_{B}^{2}}\sum_{m=0}^{+\infty}
 \frac{1}{\sqrt{2 \pi} \Gamma} e^{-(\varepsilon-\varepsilon_{m})^{2}/2 \Gamma^{2}}.
\end{equation}
Using the Poisson summation formula
\begin{equation}
\sum_{m=0}^{+\infty} f(n) = \sum_{l=-\infty}^{+\infty}
\int_{0}^{+\infty} f(t) e^{-2 \pi i lt} dt
,
\end{equation}
we can express the density of states (per spin) in the form
\begin{equation}
g(\varepsilon)=g_{0}
\left\{
1+
2
\sum_{l=1}^{+\infty}
(-1)^{l}
\cos
\left( \frac{2 \pi l \varepsilon}{\hbar \omega_{c}}
\right)
e^{-2\pi^{2}l^{2}
\left( \frac{\Gamma}{\hbar \omega_{c}}
\right)^{2}
}
\right\} \label{dos}
\end{equation}
where $g_{0}=S m^{\ast}/2\pi \hbar^{2}$ is the density of states (per spin) at zero magnetic field.
For illustration, we have plotted in Fig. \ref{dosc} the density of states obtained for three
different broadening parameters $\Gamma$.

This result (\ref{dos}) for the density of states in high magnetic fields and in the presence of a random potential with a large correlation length
has been obtained within other theoretical approaches, such as the conventional perturbation theory, \cite{Rai} the quasiclassical path integral formalism \cite{Aronov,Mirlin} or the high-field expansion. \cite{Apenko1983}
It should be noted that the obtention of a gaussian form for the Landau levels broadening in the conventional diagrammatic technique is highly nontrivial (and obtained at the price of technical difficulties, see Ref. \onlinecite{Rai}) since this technique implicitly assumes a Lorentzian broadening of the Landau levels. On the other hand, the Gaussian broadening is naturally derived from the high-field expansions \cite{Apenko1983} and the quasi-classical path integral technique. \cite{Aronov,Mirlin}

\subsection{Oscillations of the chemical potential}

One great advantage of formulating the high-field expansion in terms of real-time Green functions is the insight into the quantum
mechanical transport properties in high magnetic fields (see next Section). To
study the dependence of the transport coefficients on the magnetic field, it is
important to determine first the dependence of the chemical potential itself on the magnetic field (the location of the chemical potential with respect to the Landau levels peaks when varying the magnetic field is crucial to understand the width of the plateaus of the Hall conductance in the integer quantum Hall
effect, see Section~\ref{transport}).
For this purpose we follow the derivation of Refs. \onlinecite{Cha2001a} and \onlinecite{Cha2001} by calculating first the grand canonical thermodynamic potential $\Omega$

\begin{equation}
\Omega=-2 T \int_{0}^{+ \infty} d \varepsilon \, g(\varepsilon) \ln \left(1 + e^{\frac{\mu-\varepsilon}{T}} \right).
\end{equation}
Here the factor 2 accounts for the spin.
Inserting the expression (\ref{dos}) for the density of states, and performing the integration over the energy by using an integration by parts (for details see Ref. \onlinecite{Cha2001a}), we get that at temperatures $T \ll \mu$
\begin{equation}
\Omega=\Omega_{0}+\tilde{\Omega}
\end{equation}
where the field non-oscillating part of the thermodynamical potential is
\begin{equation}
\Omega_{0}=-g_{0} \mu^{2}
\end{equation}
and the oscillating part is
\begin{equation}
\tilde{\Omega}=g_{0} \frac{\hbar^{2}\omega_{c}^{2}}{\pi^{2}} \sum_{l=1}^{+\infty} \frac{(-1)^{l}}{l^{2}} \frac{\lambda_{l}}{\sinh \lambda_{l}}\cos\left(\frac{2 \pi l \mu }{\hbar \omega_{c}} \right) e^{-2 \pi^{2} l^{2} \left(\frac{\Gamma}{\hbar \omega_{c}} \right)^{2}}
\end{equation}
with $\lambda_{l}= 2 \pi^{2} l T/\hbar \omega_{c}$.
The number of electrons is expressed as
\begin{equation}
N=-\left(\frac{\partial \Omega}{\partial \mu} \right)_{T,B}
\end{equation}
which yields
\begin{eqnarray}
N&=& 2g_{0} \frac{\hbar \omega_{c}}{\pi}
 \sum_{l=1}^{+\infty} \frac{(-1)^{l}}{l} \frac{\lambda_{l}}{\sinh \lambda_{l}}\sin\left(\frac{2 \pi l \mu }{\hbar \omega_{c}} \right) e^{-2 \pi^{2} l^{2} \left(\frac{\Gamma}{\hbar \omega_{c}} \right)^{2}} \nonumber \\
&&
+2g_{0} \mu.
\end{eqnarray}
This equation determines the number of particles $N$ as a function of $(B,T)$ at fixed $\mu$. On the other hand, we can consider it in the thermodynamical limit as the equation for the chemical potential as the function of $(B,T)$ at a fixed number of particles $N$. Introducing the Fermi energy $\varepsilon_{F}=N/2g_{0}$,
the dependence of the chemical potential $\mu$ on the temperature $T$ and the magnetic field $B$ has then to be found from the self-consistent equation
\begin{equation}
\mu=\varepsilon_{F}-\frac{\hbar \omega_{c}}{\pi}
\sum_{l=1}^{+\infty} \frac{(-1)^{l}}{l} \sin\left(\frac{ 2 \pi l\mu}{\hbar \omega_{c}} \right) \frac{\lambda_{l}}{\sinh \lambda_{l}}
e^{-2\pi^{2}l^{2}
\left( \frac{\Gamma}{\hbar \omega_{c}}
\right)^{2}
}. \label{chem}
\end{equation}

Here  the parameter $\Gamma$, which is independent of the magnetic field, characterizes the global variations of the potential amplitudes in the bulk.
For systems with dimensions exceeding the magnetic length $l_{B}$ by several order of magnitudes the ratio $\Gamma/\hbar \omega_{c} $ can in principle take values bigger than unity if the magnetic field is not too high or/and if the system is sufficiently disordered (so that the spread of the potential amplitudes exceeds the cyclotron energy). In this case, there is a significant density between the Landau levels and we see from Eq. (\ref{chem})
that the chemical potential $\mu$ depends weakly on the magnetic field as
\begin{equation}
\mu \approx \varepsilon_{F}+\frac{\hbar \omega_{c}}{\pi}
 \sin\left(\frac{ 2 \pi \varepsilon_{F}}{\hbar \omega_{c}} \right) \frac{\lambda_{1}}{\sinh \lambda_{1}}
e^{-2\pi^{2}
\left( \frac{\Gamma}{\hbar \omega_{c}}
\right)^{2}
}.
\end{equation}
In the opposite regime $\Gamma \ll \hbar \omega_{c}$, the density of states consists of very sharp bands peaked on the Landau levels, and Eq. (\ref{chem}) yields that at low temperatures the chemical potential is pinned to a value close to $\varepsilon_{n}$  for all magnetic fields except for the integral values of  $\varepsilon_{F}/\hbar \omega_{c}$ where it experiences a sharp jump.

\begin{figure}
\includegraphics[width=\columnwidth]{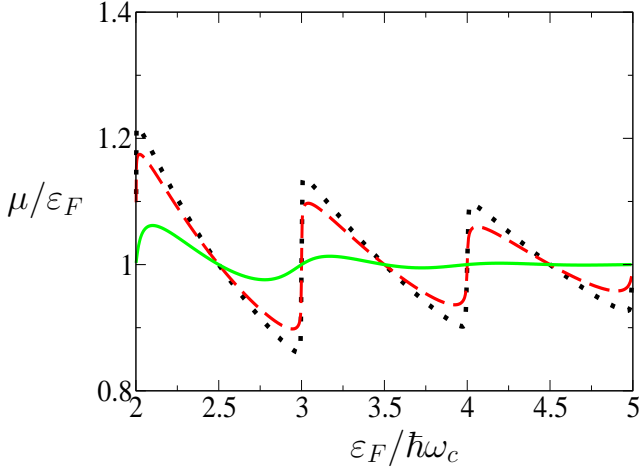}
\caption{ (color online)
Oscillations of the chemical potential as a function of the inverse magnetic field (through the ratio $\varepsilon_{F}/\hbar \omega_{c}$) for various disorder broadening parameters $\Gamma$ (same values as in Fig. \ref{dosc}; we have used the same convention for the correspondence between the curves and the parameters). The plot is made for the temperature $T=\varepsilon_{F}/200$. For large broadening, the oscillations are damped.}
\label{mu}
\end{figure}

To confirm these trends  at a more quantitative level, we have solved numerically the self-consistent equation (\ref{chem}).
In Fig. \ref{mu}, we have represented the oscillations of the chemical potential
as a function of the inverse magnetic field for three different values of the
disorder parameter $\Gamma$ (for convenience, we have used the same values for
$\Gamma$ as in the plot of the density of states, Fig. \ref{dosc}). The
characteristic sawtooth variation (dotted and dashed curves) is obtained for a
vanishing density of states between the Landau levels. On the other hand, the oscillations of the chemical potential $\mu$ around its zero field value ($\varepsilon_{F}$)  are smoothed and damped in the case of a significant overlap between the Landau levels.

\section{Quantum transport properties}
\label{transport}

\subsection{Derivation of the current}

We consider the problem of electrons confined in a Hall bar of finite width $L_{y}$. The edges are located at the positions $Y= \pm L_{y}/2$ and are included in our formalism by a rigid confinement brought about by infinitely high walls.
In experimental devices, a current $I_{x}$ is applied along the longitudinal $x$ direction and induces a transverse voltage difference $U_{H}$ (Hall voltage). The hamiltonian considered for the bulk of the sample is given by (\ref{Hamil}) where the potential term $V({\bf r})$ 
includes both the impurity potential and the electrostatic potential (we remind that the explicit form of the potential is not needed: we just assume that $V$ varies smoothly on the scale of the magnetic length).

We determine the total current ${\bf I}$ from the bulk Green functions. It is expressed as
\begin{eqnarray}
{\bf I} = - \frac{i}{L_{x}} \sum_{\nu,\nu'} \int \frac{d \omega}{2 \pi}G^{<}_{\nu,\nu'}(\omega) \, \langle \nu' | \hat{{\bf j}} | \nu \rangle \label{Itot}
\end{eqnarray}
where $L_{x}$ is the longitudinal length of the sample. The  matrix elements of the current density operator $\hat{{\bf j}}$ are given by (see Appendix C)
\begin{eqnarray}
\langle \nu' | \hat{{\bf j}} | \nu \rangle=\frac{ie \hbar \langle {\bf R}' | {\bf R} \rangle}{\sqrt{2} m^{\ast} l_{B}}  \left(
\begin{array}{c}
\sqrt{m} \, \delta_{m,m'+1} - \sqrt{m'} \, \delta_{m',m+1} \\
i\sqrt{m} \, \delta_{m,m'+1} + i\sqrt{m'} \, \delta_{m',m+1}
\end{array}
\right), \label{jmat}
\end{eqnarray}
and
\begin{equation}
G^{<}=\frac{1}{2} \left( G^{K}-(G^{R}-G^{A}) \right). \label{Gless}
\end{equation}
We have established in Appendix B that the identity
\begin{equation}
 G^{K}=\tanh(\omega/2T) (G^{R}-G^{A}) \label{GK}
\end{equation}
that relates the Keldysh Green function to the retarded and advanced Green functions at equilibrium is also valid at any order of the expansion theory in the stationary regime. This result corresponds to the realization of a nonequilibrium regime with a local (hydrodynamic) equilibrium.
Therefore
\begin{equation}
G^{<}=- n_{F}(\omega)\left(G^{R}-G^{A}\right). \label{Glessbis}
\end{equation}

It is straightforward to see that the leading order Green functions $G^{(0)}$ which are diagonal with respect to the quantum number $m$ do not contribute to the current since this latter involves only terms that couple the quantum numbers $m'$ and $m \pm 1$.
As emphasized in the Introduction, it is thus necessary to calculate the first contribution introducing Landau levels mixing in order to determine the quantum transport properties, i.e. it is required to work beyond  the semi-classical limit $B=\infty$ (unlike spectral properties that can be already obtained in the absence of Landau levels mixing).
Then, inserting the first order Green functions $G^{(1)}$, and performing the sums over $\omega$, ${\bf R}'$ and $m'$, we get the simple expression

\begin{eqnarray}
{\bf I}=\frac{e}{h}  \sum_{m=0}^{+ \infty} \int \!\! \frac{d^{2} {\bf R}}{L_{x}} \, n_{F}\left(\xi_{m}+V({\bf R})\right) \,
{\bm \nabla} V ({\bf R})  \times \hat{{\bf z}} .\label{simple}
\end{eqnarray}

Before deriving the quantization of the Hall conductance, let us comment on this expression. We see that a current exists only in the presence of gradients of the potential $V$.
By disregarding the quantization of the cyclotron motion and therefore the sum over Landau levels, we obtain a formula for the current that is similar to
the formula that would be obtained classically  by solving the Boltzmann equation together with the equations of motions in the limit of infinite magnetic fields (see Ref. \onlinecite{Apenko1985}). In the limit $B \to \infty$ the current is described only in terms of the drift of the center of the cyclotron orbit (in this limit the positions of the center and of the electron are in fact the same since the cyclotron radius vanishes).
In our quantum-mechanical picture, a current is obviously associated with a drift of the vortex location only. The contribution of the orbital currents (induced by the magnetic field) to the total current is taken into account through the sum over the integral quantum numbers $m$ and the quantization of the spectrum into Landau levels.
Due to this quantization of the phase circulation around the vortex that confers in a way a quantum robustness to the cyclotron motion, the above current formula holds at finite magnetic fields in any potential that varies smoothly at the scale of the magnetic length.
We propose now a transparent derivation of the quantization of the Hall conductance that differs from that presented in the Ref. \onlinecite{Apenko1985} (although the starting formula (\ref{simple}) is similar).

The crucial and remarkable point is that from the formula (\ref{simple}) we can formally extract a {\em vortex} current density ${\bf j}({\bf R})$ (we remind that here ${\bf R}$ corresponds to the vortex position, and not to the electron position)

\begin{equation}
{\bf j}({\bf R})= \frac{e}{h} \sum_{m=0}^{+\infty} n_{F}\left(\xi_{m}+V({\bf R})\right) \,
{\bm \nabla} V ({\bf R})  \times \hat{{\bf z}},
\end{equation}
which can be rewritten as
\begin{eqnarray}
{\bf j}({\bf R})=\frac{e}{h} \sum_{m=0}^{+\infty}
{\bm \nabla} \left[ N_{F}\left(\xi_{m}+V({\bf R})\right) \right] \times  \hat{{\bf z}} \label{curdens}
\end{eqnarray}
where
\begin{equation}
N_{F}(\varepsilon)=\varepsilon+T \ln n_{F}(\varepsilon)
\end{equation}
 is an antiderivative of the Fermi-Dirac function $n_{F}(\varepsilon)$. We can easily check that the current density (\ref{curdens}) obeys the continuity equation ${\bm \nabla} \cdot {\bf j}=0$.

The current flowing in the longitudinal direction is then given by
\begin{eqnarray}
I_{x}=\frac{e}{h}\int \!\! dY   \sum_{m=0}^{+ \infty}
\partial_{Y} N_{F}\left(\xi_{m}+V({\bf R})\right).
\end{eqnarray}
We can readily integrate this expression with respect to the coordinate $Y$ (which is delimited by $|Y| < L_{y}/2$ due to the rigid confinement) so that the current becomes
\begin{eqnarray}
I_{x}=\frac{e}{h} \sum_{m=0}^{+ \infty}
\left[ N_{F}\left(\xi_{m}+V_{+}\right) - N_{F}\left(\xi_{m}+V_{-}\right)\right]
\label{cur}
\end{eqnarray}
with $V_{\pm}=V(X,\pm L_{y}/2)$.
Remarkably it turns out from this integration that the bulk states do not formally contribute to the current. This property is a consequence of the continuity equation obeyed by the vortex current density in two-dimensions.
Furthermore, we see that the opposite edges (+ and -) contribute to the current with an opposite sign. This sign confers a chirality on the edges. Here, we recover from a bulk approach the importance of the edges for the transport properties in high magnetic fields that had been pointed out by several authors. \cite{Halperin,MacDonald,Buttiker}

Besides, from boundary requirements, we have to impose the condition that the current density flowing in the $y$ direction vanishes at the edges
\begin{equation}
j_{y}=-\frac{e}{h} \sum_{m=0}^{+ \infty} n_{F}(\xi_{m}+V({\bf R})) \,
\partial_{X} V_{\pm}({\bf R})=0. \label{condi}
\end{equation}
 We obtain from this boundary condition that the edges $Y=\pm L_{y}/2$ are at constant potentials ($\partial_{X} V_{\pm}=0$). Let us take  e.g. $V_{\pm} =\mp e U_{H}/2$ where $U_{H}$ is the Hall voltage. Obviously, the longitudinal conductance $G_{xx}$ is then zero.
Differentiating the current expression (\ref{cur}) with respect to the voltage $U_{H}$, we obtain the differential Hall conductance (defined by $G_{yx}=-dI/dU_{H}$)
\begin{eqnarray}
G_{yx} & = & \frac{e^{2}}{2h} \sum_{m=0}^{+ \infty} \left[
 n_{F} \left(\xi_{m}+V_{+}\right) +n_{F} \left(\xi_{m}+V_{-}\right)
\right] . \label{nl}
\end{eqnarray}
In the linear response $U_{H} \to 0$, we get straightforwardly
\begin{eqnarray}
G_{yx} & = & \frac{e^{2}}{h} \sum_{m=0}^{+ \infty}
 n_{F} (\varepsilon_{m}-\mu) , \label{Hallformula}
\\
G_{xx} & = & 0.
\end{eqnarray}
At zero temperature, the Hall conductance is therefore quantized as $G_{yx}= n e^{2}/h$ as long as the chemical potential $\varepsilon_{n} < \mu < \varepsilon_{n+1}$. This quantization is clearly associated with a vanishing longitudinal conductance  $G_{xx}=0$. 

As shown in the previous Section, the chemical potential $\mu$ can effectively take values intermediate between the Landau levels $\varepsilon_{n}$ when varying the magnetic field because of the broadening of the density of states caused by the distribution of amplitudes of the impurity potential $V$ in the bulk. 
In the non-linear regime ($U_{H} \neq 0$ in the right-hand side of the equation (\ref{nl})), we see that the Hall conductance can still be quantized at zero temperature as long as both edge states are filled or empty, i.e. when the chemical potential $\mu$ lies above or below the pair of edges levels.

It is worth stressing here the difference with the spectral arguments mentioned in the Introduction that are usually provided to explain the absence of dissipation in high magnetic fields. 
We see from the present microscopical quantum derivation of transport properties that the key result leading to the vanishing of $G_{xx}$ and to the quantization of $G_{xy}$ is the expression of the total current in terms of the edge states only.
Such a property has straightforwardly resulted from the possibility to define a vortex current density at high but finite magnetic fields.
We thus deduce that the existence of electronic vortex states, rather than the presence of a disordered potential, is responsible for the high-field localization properties. In fact, the bulk  disordered potential plays a role only at the thermodynamical level through the oscillations of the chemical potential with the magnetic field, as will be developed now.

\subsection{Study of the width of Hall plateaus}

\begin{figure}[t]
\includegraphics[width=\columnwidth]{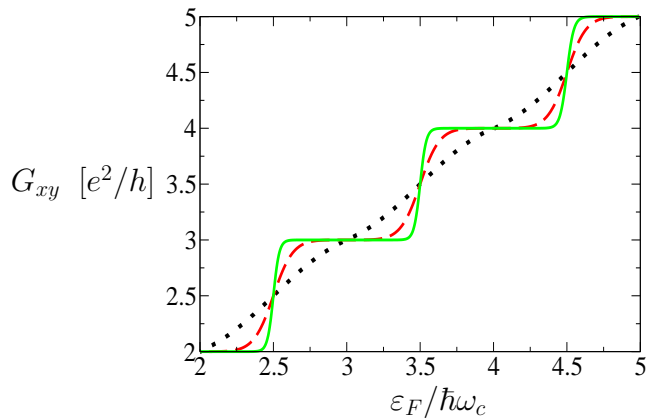}
\caption{ (color online)
Hall conductance as a function of the inverse magnetic field for various disorder broadenings. Here the temperature is  $T=\varepsilon_{F}/200$. The three disorder parameters $\Gamma$ considered are the same as in the Figs. \ref{dosc} and \ref{mu} (we have followed the same convention for the curves). The nicest plateaus are obtained for the largest broadening parameter.}
\label{HallGamma}
\end{figure}

To study the dependence of the plateaus widths on the temperature or the disorder broadening, it is crucial to take into account the dependence of the chemical potential on the magnetic field. For this purpose we have computed numerically the Hall conductance given by the formula (\ref{Hallformula}) including the self-consistent solution for the  chemical potential determined from the equation (\ref{chem}).

For a given finite temperature, the plateaus become narrower when decreasing the impurity broadening parameter, as can be clearly seen in Fig. \ref{HallGamma}.
This paradoxal trend is well known experimentally and led to the conclusion that the dirtiness of the sample is crucial for the integer quantum Hall effect. Besides, the quantum Hall effect does not occur in samples with strong disorder. The important condition to understand this limitation by the disorder is the condition for the high-field expansion. The expansion is roughly valid provided that the characteristic length of variation of the impurity potential is typically bigger than the magnetic length. This condition is fulfilled only for relatively high magnetic fields and smooth disordered potential. We note in Fig.  \ref{HallGamma} that on the other hand the disordered potential has to provide a sufficient broadening of the density of states in order to get wide plateaus.

\begin{figure}
\includegraphics[width=\columnwidth]{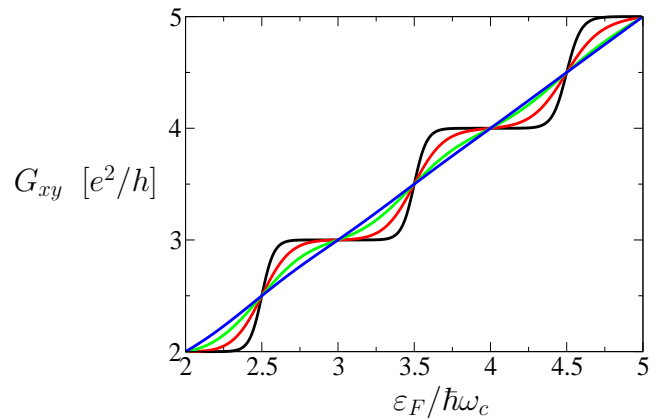}
\caption{ (color online)
Hall conductance as a function of the inverse magnetic field for various
temperatures. Here the disorder broadening parameter $\Gamma=\varepsilon_{F}/10$
corresponding to a relatively flat density of states (see Fig. \ref{dosc}). The
widest Hall plateau is obtained for the lowest temperature, here
$T=\varepsilon_{F}/80$. The width of the plateau shrinks with increasing temperature. The other curves correspond to the higher temperatures $T=\varepsilon_{F}/40$,  $T=\varepsilon_{F}/20$, and $T=\varepsilon_{F}/10$.}
\label{HallT}
\end{figure}

\begin{figure}
\includegraphics[width=\columnwidth]{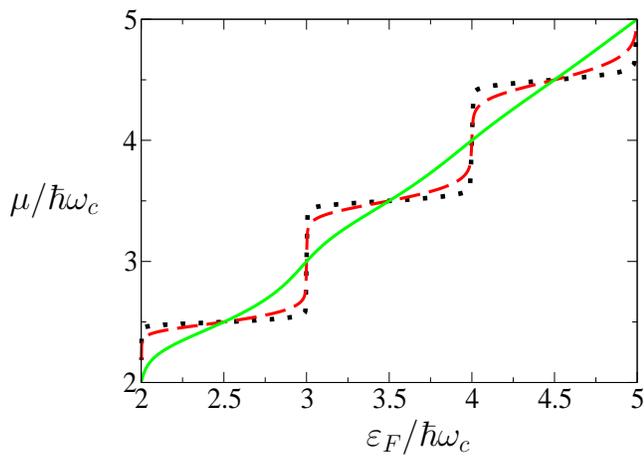}
\caption{ (color online)
The chemical potential $\mu$ over $\hbar \omega_{c}$ as a function of the magnetic field. For a small broadening of the Landau levels (dotted curve), the ratio $\mu/\hbar \omega_{c}$ remains close to the value $n+1/2$ for a wide range in magnetic fields. For a large broadening (solid curve), the chemical potential approaches the energy $(n+1/2)\hbar \omega_{c}$ only for the fields corresponding to $\varepsilon_{F}=(n+1/2)\hbar \omega_{c}$.  For the correspondence between the different curves and the chosen parameters, see Figs. \ref{dosc}, \ref{mu} and \ref{HallGamma}.}
\label{muoverB}
\end{figure}

In Fig. \ref{HallT}, we study the temperature dependence of the plateau width for a fixed disorder configuration characterized by a finite $\Gamma$.
For high temperatures, the plateaus are almost invisible and we recover the classical Hall law for the transverse conductance.
 We see that the full plateau width is restored by decreasing the temperature. This trend is found for any finite $\Gamma$, even very small.
Thus,
a large density of states between the Landau levels as seemingly often the case in experiments \cite{Eisenstein,Prange,Hadju} on the integer quantum Hall effect is not a necessary condition to observe wide plateaus of the Hall conductance. To obtain wide Hall plateaus,  a small density of states between Landau levels just requires lower temperatures.
 Nevertheless, at any finite temperature the nicest plateaus are obtained for the broadest  Landau levels.

This tendency can entirely be understood in terms of the location of the chemical potential when varying the magnetic field and thermal activation effects. For this purpose, it is instructive to plot the chemical potential $\mu$ divided by the cyclotron energy $\hbar \omega_{c}$, see
Fig. \ref{muoverB}. We note that in the quasi-absence of states between the Landau levels, the ratio $\mu/\hbar \omega_{c}$ remains very close to the value $n+1/2$ for a wide range in magnetic fields. The chemical potential is then separated from
the
energy of the edges contributing to the current by a very small energy gap, that
gives rise at finite temperatures to an important thermal activation and a
disappearance of the Hall plateaus. On the contrary, for a significant overlap between the Landau levels, the deviation of the Hall conductance from the quantized value by thermal activation effects is considerably reduced over a much wider range of magnetic fields, since the chemical potential approaches the energy of the edges more rarely.

\subsection{Higher order contributions to the current}

We have obtained the quantization of the Hall conductance by considering only the contributions of the first order Green functions to the current formula. In fact, our formalism allows to derive in a systematic and rigorous way higher order contributions.
Our calculations (the details will be published elsewhere) have revealed that the next order terms in the current formula are provided by the third-order Green functions.
As for the first order calculation, it is again possible to define a vortex current density ${\bf j}({\bf R})$. We find after cumbersome calculations that the current density due to third-order processes in Landau-levels mixing is given by
\begin{widetext}
\begin{eqnarray}
{\bf j}^{(3)}= \frac{e}{h}
\sum_{m=0}^{+\infty} 
 \frac{l_{B}^{2}}{2} 
{\bm \nabla} 
\left[
n_{F}\left(\xi_{m}+V\right) \, 
\left(
(m+1) \Delta V
+ \frac{\left|{\bm \nabla} V \right|^{2}}{\hbar \omega_{c} }
\right)
\right]
  \times \hat{{\bf z}}.
\end{eqnarray}
\end{widetext}
Obviously, this third-order contribution to the current density satisfies the continuity equation $\bm{\nabla} \cdot {\bf j}^{(3)}=0$. By following the same derivation as presented for the first process in Landau-levels mixing, we thus deduce that the total current is expressed only in terms of the edge contributions also at third-order.

In fact, from the expressions (\ref{Itot}), (\ref{jmat}) and (\ref{clom}), it is clear that it is possible to define a {\em vortex} current density at any order in the high-field expansion as
\begin{eqnarray}
{\bf j}({\bf R}) = - i \left. \sum_{m,m'} \int \frac{d \omega}{2 \pi}G^{<}_{\nu,\nu'}(\omega) \, \langle \nu' | \hat{{\bf j}} | \nu \rangle 
\right|_{{\bf R}={\bf R}'}.
\end{eqnarray}
Using the relation (\ref{Glessbis}), the vortex current density can also be written quite generally in terms of the retarded and advanced Green functions as 
\begin{eqnarray}
{\bf j}({\bf R}) =\left. \sum_{m,m'} \int \frac{d \omega}{2 \pi} i n_{F}(\omega) \left( G^{R}-G^{A}\right)_{\nu,\nu'} \, \langle \nu' | \hat{{\bf j}} | \nu \rangle 
\right|_{{\bf R}={\bf R}'}.
\end{eqnarray}
The possibility to define a vortex current density at any order in the high-field expansion has very strong implications.  Provided that the continuity equation $\bm{\nabla} \cdot {\bf j}=0$ is checked at any order (a general proof of it will be the subject of future work),
we can conclude that the expression of the total current in terms of the edges contributions only is exact as long as the high-field expansion is applicable.
We
interpret this exact result as a fundamental manifestation of the robustness of the electronic vortex states, which gives rise to strong localization features in the quantum transport properties.

\section{Discussion and conclusion}
In this Section, we would like to discuss some important issues concerning the quantum Hall effect in the light of the theory developed in the former Sections.

\subsection{Longitudinal transport}
When the Hall resistance $R_{xy}$ jumps from one Hall plateau to the adjacent one, peaks in the longitudinal resistance  $R_{xx}$ appear. \cite{vonK,Prange}
A standard picture to explain such a behavior of the longitudinal conductance is that at zero temperature the edge of the electron gas generally will follow some trajectory near the physical edge which is only on average parallel to it when the Hall conductance is in the plateau region.
As the step in the Hall conductance is approached, the equipotentials have excursions ever deeper into the bulk of the sample, and at a critical value of the Fermi energy, there is an equipotential contour that connects the two edges. This is supposed to be the origin of the peak in the diagonal conductance.


Within our present one-body approach,
we have found that the longitudinal conductance is zero. 
The problem with this zero longitudinal conductance has resulted straightforwardly from the Eq. (\ref{condi}) that expresses the absence of current flow accross the outer surfaces. To our opinion, this physical boundary condition can not be questioned. Indeed, it is the essential physical condition by which the presence of boundaries is accounted in the purely classical derivation of the Hall effect (see e.g. Ref. \onlinecite{Abri}). In a formalism making use of wavefunctions the presence of boundaries will be introduced with the condition that the wavefunction has to vanish at the edges. 
In a Green functions formalism the edges are usually taken into account via boundary conditions on the current (see e.g. Ref. \onlinecite{Nazarov} concerning the theory of mesoscopic superconductivity with quasi-classical Green functions). This is precisely the approach chosen in our theory, with the peculiarity here that the boundary conditions are applied  on the vortex current instead of the electronic current.
Such an approach assumes that there exists some degree of freedom characterizing the Green functions  that is only completely determined through the boundary conditions.
Coming back to the Dyson equation (\ref{DysonRA}) this indeterminacy of the Green functions is however hidden in the electrostatic part of the potential $V$. 

We note that the transition between Hall plateaus falls within the description of quantum phase transitions.
Clearly, the transition between the Hall plateaus is highly sensitive to any fluctuations present in the system. We remark that the explanation of the longitudinal transport by fluctuations of the electronic density has also been proposed in several works. \cite{Ruzin,Lee,Dykhne,Simon}
Recent experimental \cite{Pan2005} and theoretical \cite{Ilan2006} works have focused on the phenomenological relationship  between the diagonal and the transverse resistances observed experimentally, namely $R_{xx} \propto B \times dR_{xy}/dB$. From these works it turns out \cite{Pan2005} that all $R_{xx}$ features could be understood in terms of electron density inhomogeneities (at least in the regime of the fractional quantum Hall effect). 
Our theory has formulated the problem of transport in terms of vortex motion. However, inhomogeneities of the vortex density have not been considered in our calculations.

The above discussion clearly calls for a proper treatment of the electrostatic problem. 
By including electron-electron interactions in our formalism, the inhomogeneities of the vortex densities translate into a self-consistent determination of the local electrostatic potential (e.g. in a Hartree-Fock approximation) for which the boundary conditions on the current have to be imposed.
This can be understood as spatial variations of the local chemical potential  $\mu({\bf R})$ at local equilibrium. Clearly, according to expression (\ref{Hallformula}), such fluctuations  of the chemical potential at the edges create, when approaching the transition points between two Hall plateaus, parts of the samples characterized by a different quantization of the Hall conductance. This automatically implies that the edges are no more equipotential lines, so that there exists a longitudinal voltage difference. 
A detailed and quantitative study of this problem within our formalism is beyond the scope of the present paper and is therefore postponed for future work.

\subsection{Breakdown of the integer quantum Hall effect}

The quantum Hall effect is known \cite{Prange,Hadju} to break down at low magnetic fields and for high current intensities. These limitations can be captured qualitatively within the condition of applicability of the high magnetic fields expansion theory. Indeed, the quantum Hall effect regime holds roughly as long as the potential $V$ (which includes the electrostatic potential and the impurity potential) is smooth at the scale of the magnetic length, i.e. $l_{B} \ll \lambda$ where $\lambda$ is the typical length scale for the potential.
An increase of the current intensity induces unavoidably an increase of the net gradient of the electrostatic potential, and thus a reduction of the characteristic length scale $\lambda$ of the potential variations. Equivalently, the inequality $l_{B} \ll \lambda$ breaks down in low magnetic fields since $l_{B} \propto B^{-1/2}$ diverges when $B \to 0$.
The quantitative determination of the critical values of current and of magnetic field seems arduous since it supposes the accurate determination of the radius of convergence of the functional series (\ref{ser}) and of the high-field expansion of the Green functions.

\subsection{Relationship between the quantum Hall effect and the Shubnikov-de Haas effect}

It is worth emphasizing that the low magnetic field regime is usually characterized by another quantum effect, namely the
Shubnikov-de Haas oscillations of the longitudinal resistance $R_{xx}$, while the high magnetic field regime corresponds to the quantum Hall effect.
In the Shubnikov-de Haas regime, $R_{xx}$ presents minima between the Landau levels that become deeper and deeper when increasing the strength of the magnetic field, whereas there is a preformation of rough steps in the $R_{xy}$ features. In the quantum Hall effect regime, $R_{xx}$ drops to zero.
The passage from the Shubnikov-de Haas regime to the quantum Hall regime is relatively smooth. It
suggests the existence of a physical relationship between these two quantum effects.

The theory of the Shubnikov-de Haas effect for the two-dimensional electron gas in the presence of a disordered potential has been developed a long time ago  within the Green functions formalism and the standard impurity averaging
procedure. \cite{Ando1974,Ando1982}
The oscillations of $R_{xx}$ are usually understood in terms of the oscillations of the density of states. Within this picture, the minima of $R_{xx}$ are related to the minima of the density of states.
Theoretical derivations \cite{Richter1995,Hackenbroich1995a,Hackenbroich1995b,Hackenbroich1996,Mirlin} of these oscillations making use of semi-classical Green functions provide a deeper interpretation concerning the physical origin of the Shubnikov-de Haas effect. Within this semi-classical approach, the oscillations of $R_{xx}$ and of $R_{xy}$ are interpreted as interference effects due to a fraction of electrons following classical periodic orbits, in close relation to the Gutzwiller's trace formula \cite{Gutzwiller} for the density of states. \cite{note2}
 This fraction $P=\exp(-2 \pi R_{c}/l)$ (where $l$ is the mean free path) of electrons participating to magnetic quantum effects in low fields  \cite{note1} corresponds to the classical probability that the electron will complete a cyclotron orbit without having suffered a collision. \cite{Bobylev1995}
Note that the disorder tends to reduce this probability and thus destroys these magnetic interference effects. This is in complete contrast with the weak localization effects that are induced by the disorder and are suppressed by the application of a weak magnetic field (see e.g. Ref. \onlinecite{Altshuler}).
Since the cyclotron radius diminishes with the magnetic field, the fraction of circling electrons increases and the minima of $R_{xx}$ become more and more pronounced. In the limit of infinite magnetic field, we expect from the semi-classical picture that all the electrons participate to these magnetic interference effects.

In the present work, we have shown that  the quantization of the orbital degree of freedom confers a robustness on the periodic cyclotron motion, so that the electronic states have a probability higher than the classical one (given above) of returning to a same given position. The ineffective energy exchange between the orbital degree of freedom and the vortex degree of freedom occurs consequently in any local region of smooth potential in high but finite magnetic fields, i.e. not only in the semi-classical limit $B=\infty$. This property resulting from the topological character of the circulation quantum number $m$ is illustrated by the possibility to define a vortex current density $j({\bf R})$ at any order of the high magnetic fields gradient expansion.
The electronic system enters into the quantum Hall regime   when all the electronic vortex states contribute to the magnetic interference effects (strong localization).
Accordingly, the Shubnikov-de Haas effect can effectively  be seen as a precursor to the quantum Hall effect.

\subsection{Conclusion}

The insensitivity to the type of impurities and the reproductivity of the experimental results  independently of the type of host material has suggested that the quantum Hall effect is due to a fundamental principle.
In the first half of the paper, we have shown that the quantization of the energy spectrum into Landau levels is associated with the topological condition of the single-valuedness of the single-particle wavefunction. Electronic states can thus be characterized by vortex states whose nature comes out in the presence of a perturbation (for instance a disordered potential). As developed in the second half of the paper, the robustness of the vortex states is responsible for the remarkable quantum transport properties of a two-dimensional electron gas  in high {\em finite} magnetic fields, namely the vanishing of the longitudinal conductance and the quantization of the Hall conductance.

By reformulating the high-field expansion theory using a vortex states representation, we have rigourously derived within the formalism of real-time Green functions a quantum mechanical expression for the current. Within our vortex representation, a  current is obviously associated with a (hydrodynamic) vortex flow. We have shown that this quantum mechanical property is an exact result that holds as long as the high-field expansion is applicable.
We have demonstrated explicitly from a bulk derivation up to third order processes in Landau-levels mixing that the total current can be expressed entirely in terms of edge contributions. This remarkable property  together with boundary requirements leads directly to the absence of dissipation and to the quantization of the Hall conductance.

We have emphasized that the density of states is not necessarily small between the Landau levels in the vortex regime. By computing the oscillations of the chemical potential with the magnetic field, we have pointed out that at any finite temperature the widest widths for the Hall plateaus are precisely obtained in the case of the broadest Landau levels, in agreement with experimental observations.

The quantum mechanical picture and the formalism developed in the paper are appealing to explain several issues concerning the quantum Hall effect. We have already pointed out qualitatively the relationship between the Shubnikov-de Haas regime in low magnetic fields and the quantum Hall regime in higher magnetic fields. Moreover, as also underlined, our formalism allows a systematic calculation of higher order contributions to the current.
Finally, we hope that it will be possible to generalize the presented approach of the quantum transport to incorporate the interaction effects between the electrons that appear to be crucial to capture the dissipative longitudinal transport.

\section*{Acknowledgments}

Stimulating discussions with M. Houzet, T. L\"{o}fwander, V.P. Mineev, M. Pletyukhov, and A. Poenicke are gratefully acknowledged.

\begin{widetext}

\appendix

\section{Overlap}

In this Appendix, we calculate the overlap between the two vortex states $| m, {\bf R} \rangle$ and
$| m', {\bf R}' \rangle$
\begin{eqnarray}
 \langle m, {\bf R} | m', {\bf R}'\rangle=\int d^{2}{\bf r} \, \Psi_{m, {\bf R}}^{\ast}({\bf r}) \Psi_{m', {\bf R}'} ({\bf r}).
\end{eqnarray}
Making first the change of variables
${\bf u}={\bf r}-({\bf R}+{\bf R}')/2
$ and
 using the cartesian coordinates ${\bf u}=(x,y)$, we obtain
\begin{eqnarray}
 \langle m, {\bf R} |m', {\bf R}' \rangle= C_{m}C_{m'}\int \frac{d^{2}{\bf u}}{l_{B}^{2}}
\left[\frac{(x+d_{x})-i (y+d_{y})}{\sqrt{2}l_{B}}
\right]^{m} \left[\frac{(x-d_{x})+i (y-d_{y})}{\sqrt{2} l_{B}}
\right]^{m'}
 \, e^{-\frac{(x-i d_{y})^{2}+(y+i d_{x})^{2}}{2 l_{B}^{2}}} \nonumber \\
\times
e^{-\frac{({\bf R}-{\bf R}')^{2}-2 i \hat{{\bf z}}\cdot ({\bf R} \times {\bf R}')}{4l_{B}^{2}}}
\end{eqnarray}
with ${\bf d}=({\bf R}'-{\bf R})/2$.
We then make the shift of  the complex arguments $id_{y}$ and $i d_{x}$ in the $x$ and $y$ integrations respectively.
Because the functions are analytic it is possible to deform the contour to the real axis so that
\begin{eqnarray}
 \langle m, {\bf R} |m', {\bf R}' \rangle= C_{m}C_{m'}\int \frac{d^{2}{\bf u}}{l_{B}^{2}}
\left[\frac{x-i y}{\sqrt{2}l_{B}}
\right]^{m} \left[\frac{x+i y}{\sqrt{2} l_{B}}
\right]^{m'}
 \, e^{-\frac{x^{2}+y^{2}}{2 l_{B}^{2}}} \,
e^{-\frac{({\bf R}-{\bf R}')^{2}-2 i \hat{{\bf z}}\cdot ({\bf R} \times {\bf R}')}{4l_{B}^{2}}}.
\end{eqnarray}
The remaining integrations yield the result
\begin{eqnarray}
 \langle m, {\bf R} | m', {\bf R}' \rangle=
\exp\left
[-\frac{({\bf R}-{\bf R}')^{2}}{4 l_{B}^{2}} \right]\exp\left[\frac{i \hat{{\bf z}} \cdot ({\bf R} \times {\bf R}')}{2 l_{B}^{2}}
\right]\, \delta_{m \,m'}.
\end{eqnarray}

The overlap contains a phase factor resulting from the interference effects between the states $\left|m,{\bf R} \right. \rangle $ and $\left| m,{\bf R}' \right. \rangle$.
The basis of vortex states becomes orthogonal with respect to the quantum number ${\bf R}$ only in the limit of infinite magnetic fields $l_{B} \to 0$.

In fact, from a  superposition of the states $\left| m,{\bf R} \right. \rangle $ with different vortex positions, it is possible to construct  orthogonal sets of eigenstates at finite magnetic fields.
For example, the states
\begin{equation}
\Psi_{m,X}=\int_{- \infty}^{+\infty} d Y  \Psi_{m, {\bf R} } \label{pY}
\end{equation}
obtained  from the superposition of states  $\left| m,{\bf R} \right. \rangle $   (expressed in the symmetrical gauge) with their vortex positions on the axis $Y$ form an orthogonal set. Inserting the expression (\ref{vortex}) of the vortex states we perform the integration over $Y$ in Eq. (\ref{pY}) using the
identity
\begin{eqnarray}
\int_{-\infty}^{+ \infty} X^{m} \exp\left[- (X-\beta)^{2}\right]=\frac{\sqrt{\pi}}{(2i)^{n}} H_{m}(i \beta)
\end{eqnarray}
where $H_{m}$ is the Hermite polynomial of degree $m$. After calculation,  we find that the states (\ref{pY}) can be expressed as
\begin{eqnarray}
\Psi_{m,X}({\bf r})= \frac{ e^{-\frac{X^{2}}{8 l_{B}^{2}}}}{m! (\sqrt{2})^{m-1}} \, e^{-\frac{iyX}{2 l_{B}^{2}}} \, \exp \left[-\frac{(x-X/2)^{2}}{4l_{B}^{2}}\right]  H_{m}\left(\frac{x-X/2}{l_{B}} \right)
,
\label{Landausym}
\end{eqnarray}
which is nothing else than the expression of the (non normalized) Landau states (\ref{trans}) written in the symmetrical gauge. From the comparison between the expressions (\ref{Landausym}) and (\ref{trans}), we can identify $p=-X/2 l_{B}^{2}$ and $x_{0}=X/2$.

\section{Derivation of the Keldysh Green function}
In this Appendix we derive the equations obeyed by the Green functions in the Keldysh space within the vortex states representation. The quantum kinetic equation for the Keldysh Green function is then solved within the high magnetic field expansion theory.

\subsection{Dyson equation in the Keldysh space}

The Green functions in presence of the perturbation $V$ are obtained from the Dyson equation
\begin{equation}
\tilde{G}=\tilde{G}_{0}+\tilde{G}_{0} V \tilde{G}
\end{equation}
where $\tilde{G}$ designates the 2 x 2 matrix in the Keldysh space defined as

\begin{equation}
\tilde{G}=\left(
\begin{array}{cc}
G^{R} & G^{K} \\
0 & G^{A}
\end{array}
\right).
\end{equation}
Here $V$ may contain a disordered potential and an electrostatic potential.
Inserting the quantum numbers $|\nu \rangle =|m,{\bf R} \rangle$ and considering that $V$ is local in time, we have
\begin{equation}
\tilde{G}(\nu_{1},t_{1};\nu_{2},t_{2})=\tilde{G}_{0}(\nu_{1},t_{1};\nu_{2},t_{2})+ \sum_{\nu_{3},\nu_{4}} \int \!\! dt_{3} \tilde{G}_{0}(\nu_{1},t_{1};\nu_{3},t_{3}) V_{\nu_{3},\nu_{4}}  \tilde{G}(\nu_{4},t_{3};\nu_{2},t_{2}).
\end{equation}
By applying the time derivative to the Dyson equation, we thus obtain the system of equations for the three Green functions
\begin{eqnarray}
\left(
i \frac{\partial}{\partial t_{1}}- \xi_{m_{1}} \right) G^{R,A}(\nu_{1},t_{1};\nu_{2},t_{2})&=&\langle \nu_{1} | \nu_{2}\rangle \delta(t_{1}-t_{2}) + \sum_{\nu_{3}}  V_{\nu_{1}, \nu_{3}}  G^{R,A}(\nu_{3},t_{1};\nu_{2},t_{2})  ,\\
\left(
i \frac{\partial}{\partial t_{1}}- \xi_{m_{1}} \right) G^{K}(\nu_{1},t_{1};\nu_{2},t_{2})& = & \sum_{\nu_{3}} V_{\nu_{1}, \nu_{3}}  G^{K}(\nu_{3},t_{1};\nu_{2},t_{2})  \label{K1},
\end{eqnarray}
where we have used the equation of motion for $\tilde{G}_{0}$
\begin{equation}
\left(
i \frac{\partial}{\partial t_{1}}- \xi_{m_{1}} \right) \tilde{G}_{0}(\nu_{1},t_{1};\nu_{2},t_{2})=\langle \nu_{1} | \nu_{2} \rangle \delta(t_{1}-t_{2}) \tilde{I}
\end{equation}
with $\tilde{I}$ the unit matrix in the Keldysh space.
From the other Dyson equation ($\tilde{G}=\tilde{G}_{0}+\tilde{G} V \tilde{G}_{0}$) and by applying the derivation with respect to the time $t_{2}$, we get another equation for the Keldysh Green function
\begin{eqnarray}
\left(
-i \frac{\partial}{\partial t_{2}}- \xi_{m_{2}} \right) G^{K}(\nu_{1},t_{1};\nu_{2},t_{2})=  \sum_{\nu_{3}} G^{K}(\nu_{1},t_{1};\nu_{3},t_{2})V_{\nu_{3}, \nu_{2}} . \label{K2}
\end{eqnarray}

It is direct to see that the retarded and advanced Green functions depend only on the time difference $\tau=t_{1}-t_{2}$. Fourier transforming with respect to $t_{1}-t_{2}$, we get the equations
\begin{equation}
(\omega-\xi_{m_{1}}\pm i \delta) \, G^{R,A}_{\nu_{1},\nu_{2}}(\omega)= \langle \nu_{1} |\nu_{2} \rangle+\sum_{\nu_{3}} V_{\nu_{1},\nu_{3}}G^{R,A}_{\nu_{3},\nu_{2}}(\omega).
\end{equation}
As for the Keldysh component, we take the sum and the difference of Eqs. (\ref{K1})-(\ref{K2}), introduce the relative time $\tau=t_{1}-t_{2}$ and the time ${\cal T}=(t_{1}+t_{2})/2$,  and Fourier transform with respect to $\tau$, which yields
\begin{eqnarray}
\left(2\omega-[\xi_{m_{1}}+\xi_{m_{2}}]\right) \, G^{K}_{\nu_{1},\nu_{2}}({\cal T},\omega)= \sum_{\nu_{3}} \left[ V_{\nu_{1},\nu_{3}}G^{K}_{\nu_{3},\nu_{2}}({\cal T},\omega)+ G^{K}_{\nu_{1},\nu_{3}}({\cal T},\omega)  V_{\nu_{3},\nu_{2}} \right], \label{Kespectral}\\
\left(  i\frac{\partial}{\partial {\cal T}}-[\xi_{m_{1}}-\xi_{m_{2}}] \right) \, G^{K}_{\nu_{1},\nu_{2}}({\cal T},\omega)= \sum_{\nu_{3}} \left[ V_{\nu_{1},\nu_{3}}G^{K}_{\nu_{3},\nu_{2}}({\cal T},\omega)- G^{K}_{\nu_{1},\nu_{3}}({\cal T},\omega)  V_{\nu_{3},\nu_{2}} \right]. \label{Kedistribution}
\end{eqnarray}

\subsection{Derivation of  Keldysh Green function in the high magnetic field expansion}

As for the retarded and advanced Green functions in Sec. III.C, the Keldysh Green function is searched under the form of an expansion in terms of the order of $(l_{B}/\lambda)^{q}$. The leading order contribution obeys the equation

\begin{eqnarray}
\left(2\omega-[\xi_{m_{1}}+\xi_{m_{2}}]\right) \, G^{(0) \, K}_{\nu_{1},\nu_{2}}({\cal T},\omega)= \sum_{\nu_{3}} \left[ V_{\nu_{1},\nu_{3}}^{(0)} \langle
{\bf R}_{1}|{\bf R}_{3}\rangle
G^{(0) \, K}_{\nu_{3},\nu_{2}}({\cal T},\omega)+ G^{(0) \, K}_{\nu_{1},\nu_{3}}({\cal T},\omega)  V_{\nu_{3},\nu_{2}}^{(0)}
 \langle
{\bf R}_{3}|{\bf R}_{2}\rangle
 \right],\\
\left(i\frac{\partial}{\partial {\cal T}}-[\xi_{m_{1}}-\xi_{m_{2}}] \right) \, G^{(0) \, K}_{\nu_{1},\nu_{2}}({\cal T},\omega)= \sum_{\nu_{3}} \left[ V^{(0)}_{\nu_{1},\nu_{3}}
\langle
{\bf R}_{1}|{\bf R}_{3}\rangle
G^{(0) \, K}_{\nu_{3},\nu_{2}}({\cal T},\omega)- G^{(0) \, K}_{\nu_{1},\nu_{3}}({\cal T},\omega)  V^{(0)}_{\nu_{3},\nu_{2}}
\langle
{\bf R}_{3}|{\bf R}_{2}\rangle
\right].
\end{eqnarray}
For the second terms in the right-hand side of these two latter equations, we can take the potential $V$ at the point ${\bf R}_{2}$ instead of ${\bf R}_{3}$ since at leading order we disregard correcting terms of the order of $l_{B}/\lambda$. Therefore we get
\begin{eqnarray}
\left(2\omega-[\xi_{m_{1}}+\xi_{m_{2}}] - V({\bf R}_{1}) - V({\bf R}_{2})\right) \, G^{(0) \, K}_{\nu_{1},\nu_{2}}({\cal T},\omega)= 0,  \\
\left(i\frac{\partial}{\partial {\cal T}}-[\xi_{m_{1}}-\xi_{m_{2}}] - V({\bf R}_{1}) +V({\bf R}_{2})\right) \, G^{(0) \, K}_{\nu_{1},\nu_{2}}({\cal T},\omega)= 0.
\end{eqnarray}
Disregarding higher order correction terms, this
system for the zero order Keldysh Green function can in fact be written as
\begin{eqnarray}
\left(2\omega-[\xi_{m_{1}}+\xi_{m_{2}}] - 2 V({\bf R}_{1})\right) \, G^{(0) \, K}_{\nu_{1},\nu_{2}}({\cal T},\omega)= 0, \label{first} \\
\left(i\frac{\partial}{\partial {\cal T}}-[\xi_{m_{1}}-\xi_{m_{2}}] \right) \, G^{(0) \, K}_{\nu_{1},\nu_{2}}({\cal T},\omega)= 0. \label{second}
\end{eqnarray}
The
equation (\ref{first}) for the zero order Keldysh Green function yields straightforwardly the following spectral dependence
\begin{equation}
 G^{(0) \, K}_{\nu_{1},\nu_{2}}({\cal T},\omega) \propto \delta\left(\omega -[\xi_{m_{1}}+\xi_{m_{2}}]/2 -  V({\bf R}_{1})\right).
\end{equation}
The differential equation (\ref{second}) is readily integrated:
\begin{equation}
 G^{(0) \, K}_{\nu_{1},\nu_{2}}({\cal T},\omega)=C_{\nu_{1},\nu_{2}}(\omega) \, \exp\left( i [\xi_{m2}-\xi_{m1}] {\cal T}\right)
\end{equation}
where $C_{\nu_{1},\nu_{2}}(\omega)$ has to be determined from an initial condition.
Taking into account that the perturbation $V$ (including the external field $E$) is switched on adiabatically with the equilibrium initial condition at ${\cal T}=-\infty$
\begin{equation}
 G^{(0) \, K}_{\nu_{1},\nu_{2}}(- \infty,\omega)=G_{0}^{K}(\omega)=-2 i \pi \tanh\left(\omega/2T \right) \langle \nu_{1}| \nu_{2} \rangle \, \delta(\omega-\xi_{m_{1}}),
\end{equation}
we obtain
\begin{equation}
 G^{(0) \, K}_{\nu_{1},\nu_{2}}({\cal T},\omega) =- 2 i \pi \tanh\left(\omega/2T \right) \langle \nu_{1}| \nu_{2} \rangle \, \delta \left(\omega -\xi_{m_{1}} -  V({\bf R}_{1})\right)=\tanh(\omega/2T) \left( G^{(0) \, R}-G^{(0) \, A}\right).
\end{equation}
Therefore, at leading order the Keldysh Green function is still diagonal with respect to the quantum numbers $m$.

We shall determine the higher order Keldysh Green function $G^{(q) \, K}$ within the stationary regime.
Then, the first order Keldysh Green function  is determined for example from the equation resulting from the sum of Eq. (\ref{Kedistribution}) and (\ref{Kespectral}) which yields up to first order in $l_{B}/\lambda$
\begin{eqnarray}
\left( \omega -\xi_{m_{1}}-V({\bf R}_{1}) \right) \, G^{(1) \, K}_{\nu_{1},\nu_{2}}(\omega)
= \sum_{\nu_{3}} V^{(1)}_{\nu_{1},\nu_{3}}
\langle {\bf R}_{1} | {\bf R}_{3}\rangle
G^{(0) \, K}_{\nu_{3},\nu_{2}}(\omega).
\end{eqnarray}
Inserting the zero order Green function and performing the sum over $\nu_{3}$, we get
\begin{equation}
G^{(1) \, K} = \tanh(\omega/2T) \left( G^{(1) \, R}-G^{(1) \, A}\right). \label{prop}
\end{equation}
We see that the structure of the Dyson equation for the Keldysh Green function of orders higher than one  in the stationary regime is similar to  the structure of the Dyson equation for the retarded and advanced Green functions at the same order.
Consequently, we can deduce that the proportionality (\ref{prop}) between the Keldysh Green function and the retarded and advanced Green functions holds at any order of the high-magnetic field expansion theory. 

\section{ Matrix elements of the current density operator}

In this Appendix, we derive the matrix elements of the current density operator in the vortex states representation.
In the space representation ($|{\bf r} \rangle$), the current density operator is written as
\begin{equation}
\hat{{\bf j}}({\bf r})=\frac{e}{m^{\ast}}\left( \frac{\hbar}{i} {\bm \nabla}-\frac{e}{c} {\bf A} \right).
\end{equation}
Adopting the symmetrical gauge
\begin{equation}
{\bf A}=\frac{B}{2}\left( \begin{array}{c} -y \\ x\end{array}\right)=\frac{\hbar c}{|e|}\frac{1}{2 l_{B}^{2}}\left( \begin{array}{c} -y \\ x\end{array}\right),
\end{equation}
we express the matrix elements of the current density operator as
\begin{eqnarray}
\langle m,{\bf R} | \hat{{\bf j}}| m',{\bf R}'\rangle &=&
\frac{e \hbar }{m^{\ast}}
\int d^{2}{\bf r} \, \Psi_{m, {\bf R}}^{\ast}({\bf r}) \left[- i{\bm \nabla}-\frac{1}{2 l_{B}^{2}} \left( \begin{array}{c} - y \\ x \end{array}\right)  \right]
\Psi_{m', {\bf R}'} ({\bf r}) \\
&=& \frac{e \hbar}{\sqrt{2} m^{\ast}  l_{B}}\left(\begin{array}{c}
i \sqrt{m'} \,\langle m, {\bf R}| m'-1,{\bf R}'\rangle-i\sqrt{m'+1}\,\langle m, {\bf R}| m'+1,{\bf R}'\rangle\\
 \sqrt{m'} \,\langle m, {\bf R}| m'-1,{\bf R}'\rangle+\sqrt{m'+1}\,\langle m, {\bf R}| m'+1,{\bf R}'\rangle
\end{array} \right)
 \\
&=& \frac{e \hbar}{\sqrt{2} m^{\ast}  l_{B}}\left(\begin{array}{c}
i \sqrt{m'} \,\delta_{m,m'-1}-i\sqrt{m}\, \delta_{m,m'+1}\\
\sqrt{m'} \,\delta_{m,m'-1}+\sqrt{m}\, \delta_{m,m'+1}
\end{array} \right) \langle m,{\bf R}|m, {\bf R}'\rangle.
\end{eqnarray}
Note that these matrix elements  of the current density operator in the vortex states basis  couple adjacent Landau levels exactly as in  the basis of Landau states.

\end{widetext}


\begin{thebibliography}{99}

\bibitem{vonK}
K. v. Klitzing, G. Dorda, and M. Pepper, Phys. Rev. Lett. \textbf{45}, 494 (1980).


\bibitem{Prange}
{\em The Quantum Hall effect}, edited by R.E. Prange and S.M. Girvin (New York, Springer, 1987).

\bibitem{Hadju}
M. Janssen, O. Viehweger, U. Fastenrath, and J. Hadju, {\em Introduction to the Theory of the Integer Quantum Hall Effect} (VCH, Germany, 1994).

\bibitem{Huckestein} B. Huckestein, Rev. Mod. Phys. \textbf{67}, 357 (1995).

\bibitem{Laughlin}
R.B. Laughlin, Phys. Rev. B \textbf{23}, 5632 (1981).

\bibitem{Halperin}
B.I. Halperin, Phys. Rev. B \textbf{25}, 2185 (1982).

\bibitem{Streda}
P. Streda, J. Phys. C: Solid State Phys. \textbf{15}, L717 (1982).

\bibitem{Thouless}
D.J. Thouless, M. Kohmoto, M.P. Nightingale, and M. den Nijs, Phys. Rev. Lett. \textbf{49}, 405 (1982).

\bibitem{Prange2}
R.E. Prange and R. Joynt, Phys. Rev. B \textbf{25}, 2943 (1982).

\bibitem{Iordanski}
S.V. Iordansky, Solid State Commun. \textbf{43}, 1 (1982).


\bibitem{Trugman}
S.A. Trugman, Phys. Rev. B \textbf{27}, 7539 (1983).


\bibitem{Kazarinov}
R.F. Kazarinov and S. Luryi, Phys. Rev. B \textbf{25}, 7626 (1982); S. Luryi and R.F. Kazarinov, Phys. Rev. B \textbf{27}, 1386 (1983).

\bibitem{Joynt}
R. Joynt and R.E. Prange, Phys. Rev. B \textbf{29}, 3303 (1984).

\bibitem{Apenko1985}
S.M. Apenko and Yu.E. Lozovik, J. Phys. C: Solid State Phys. \textbf{18}, 1197 (1985).

\bibitem{Shapiro}
B. Shapiro, Phys. Rev. B \textbf{33}, 8447 (1986).


\bibitem{MacDonald}
A.H. MacDonald and P. Streda, Phys. Rev. B \textbf{29}, 1616 (1984).

\bibitem{Buttiker}
M. B\"{u}ttiker, Phys. Rev. B \textbf{38}, 9375 (1988).



\bibitem{Rai}
M.E. Raikh and T.V. Shahbazyan, Phys. Rev. B \textbf{47}, 1522 (1993).


\bibitem{Abrikosov}
A.A. Abrikosov, L.P. Gorkov, and I.E. Dzyaloshinski, {\em Methods of Quantum Field Theory in Statistical Physics} (Prentice Hall Inc., New Jersey, 1964).





\bibitem{Fog}
M.M. Fogler, A.Yu. Dobin, V.I. Perel, and B.I. Shklovskii, Phys. Rev. B \textbf{56}, 6823 (1997).


\bibitem{Apenko1983}
S.M. Apenko and Yu.E. Lozovik, J. Phys. C: Solid State Phys. \textbf{16}, L591 (1983).

\bibitem{Apenko1984}
S.M. Apenko and Yu.E. Lozovik, J. Phys. C: Solid State Phys. \textbf{17}, 3585 (1984).


\bibitem{Landau}
L. Landau, Z. Phys. \textbf{64}, 629 (1930).

\bibitem{Johnson}
M.H. Johnson and B.A. Lippmann, Phys. Rev. \textbf{76}, 828 (1949).

\bibitem{Madelung}
E. Madelung, Z. Phys. \textbf{40}, 322 (1926).

\bibitem{jap}

T. Takabayasi, Prog. Theor. Phys. \textbf{69}, 1323 (1983).


\bibitem{Malkin1969}
I.A. Malkin and V.I. Man'ko, Sov. Phys. JETP \textbf{28}, 527 (1969).

\bibitem{Materdey2003}
T.B. Materdey and C.E. Seyler, Int. J. of Mod. Phys. \textbf{17}, 4683 (2003).

\bibitem{Glauber1963}
R.J. Glauber, Phys. Rev. \textbf{131}, 2766 (1963).

\bibitem{Zhang1990}
W-M. Zhang, D.H. Feng, and R. Gilmore, Rev. Mod. Phys. \textbf{62}, 867 (1990).

\bibitem{Note}
Note that it is possible to construct states that are entirely coherent with respect to the two degrees of freedom
(see Refs. \onlinecite{Feldman1970} and \onlinecite{Varro1984}). Contrary to the vortex states, theses states do not have the property to be eigenstates of the hamiltonian for the problem of an electron in a magnetic field.

\bibitem{Feldman1970}
A. Feldman and and A.H. Kahn, Phys. Rev. \textbf{1}, 4584 (1970).

\bibitem{Varro1984}
S. Varro, J. Phys. A: Math. Gen. \textbf{17}, 1631 (1984).

\bibitem{Mahan}
G.D. Mahan, {\em Many-particle physics} (Plenum Press, New York, 2nd edition, 1990).

\bibitem{Rammer}
J. Rammer and H. Smith, Rev. Mod. Phys. \textbf{58}, 323 (1986).

\bibitem{Aronov}
A.G. Aronov, E. Altshuler, A.D. Mirlin, and P. W\"{o}lfle, Europhys. lett. \textbf{29}, 239 (1995).

\bibitem{Mirlin}
A.D. Mirlin, E. Altshuler, and P. W\"{o}lfle, Ann. Physik \textbf{5}, 281 (1996).

\bibitem{Cha2001a}
T. Champel and V.P. Mineev, Philos. Mag. \textbf{81}, 55 (2001).


\bibitem{Cha2001}
T. Champel, Phys. Rev. B \textbf{64}, 054407 (2001).

\bibitem{Eisenstein}
J.P. Eisenstein {\em et al.}, Phys. Rev. Lett. \textbf{55}, 875 (1985).

\bibitem{Abri}
A.A. Abrikosov, {\em Fundamentals of the theory of metals} (Elsevier, Amsterdam, 1988).

\bibitem{Nazarov} Yu.V. Nazarov, Superlattices Microstruct. \textbf{25}, 1221 (1999).


\bibitem{Ruzin}
I.M. Ruzin, Phys. Rev. B \textbf{47}, 15727 (1993).

\bibitem{Lee}
D.B. Chklovskii and P.A. Lee, Phys. Rev. B \textbf{48}, 18060 (1993).

\bibitem{Dykhne}
A.M. Dykhne and I.M. Ruzin, Phys. Rev. B \textbf{50}, 2369 (1994).

\bibitem{Simon}
S.H. Simon and B.I. Halperin, Phys. Rev. Lett. \textbf{73}, 3278 (1994).


\bibitem{Pan2005}
W. Pan {\em et al.}, Phys. Rev. Lett. \textbf{95}, 066808 (2005).

\bibitem{Ilan2006} R. Ilan, N.R. Cooper, and A. Stern, Phys. Rev. B \textbf{73}, 235333 (2006).

\bibitem{Ando1974}
T. Ando, J. Phys. Soc. Jpn. \textbf{37}, 1233 (1974).

\bibitem{Ando1982}
T. Ando, A.B. Fowler, and F. Stern, Rev. Mod. Phys. \textbf{54}, 437 (1982).

\bibitem{Richter1995}
K. Richter, Europhys. Lett. \textbf{29}, 7 (1995).

\bibitem{Hackenbroich1995a}
G. Hackenbroich and F. von Oppen, Europhys. Lett. \textbf{29}, 151 (1995).

\bibitem{Hackenbroich1995b}
G. Hackenbroich and F. von Oppen, Z. Phys.  B \textbf{97}, 157 (1995).

\bibitem{Hackenbroich1996}
G. Hackenbroich and F. von Oppen, Ann. Physik \textbf{5}, 696 (1996).


\bibitem{Gutzwiller}
M.C. Gutzwiller, {\em Chaos in Classical and Quantum Mechanics} (Springer, New York, 1990).

\bibitem{note2}
The Gutzwiller trace formula leads also to an alternative semi-classical derivation of the magnetic quantum oscillations of the magnetization (de Haas-van Alphen effect) in low magnetic fields.

\bibitem{note1}
The elastic scattering time is assumed within these semi-classical theories to be independent of the magnetic field, which in principle is valid for a two-dimensional electron gas only in low magnetic fields.

\bibitem{Bobylev1995}
A.V. Bobylev, F.A. Maao, A. Hansen, and E.H. Hauge, Phys. Rev. Lett. \textbf{75}, 197 (1995).

\bibitem{Altshuler}
B.L. Altshuler, A.G. Aronov, D.E. Khmelnitskii, and A.I. Larkin, {\em Coherent Effects in Disordered Conductors} (MIR, Moscow, 1982).



\end{thebibliography}
\end{document}